\documentclass[10pt,aps,prb,twocolumn,superscriptaddress,showpacs,floatfix]{revtex4-2}
\usepackage{color}
\usepackage{mathtools}
\usepackage[caption=false]{subfig}
\usepackage{bm}        
\usepackage{amssymb}   
\usepackage{amsmath}
\usepackage{graphicx}
\usepackage[caption=false]{subfig}
\usepackage{braket}
\usepackage{textcomp}
\usepackage{natbib}
\usepackage [latin1]{inputenc}
\usepackage{siunitx}
\usepackage{blkarray, bigstrut}
\usepackage{xcolor}
\usepackage[colorlinks=true,citecolor=blue,linkcolor=magenta]{hyperref}
\usepackage{cleveref}

\date{}
\begin{document}
\title{\textcolor{black}{\texorpdfstring{Unusual Valence of Ru and Prediction of Magnetism, Anomalous Hall Conductivity in a Newly Synthesized Double Perovskite Compound Ca$_2$CoRuO$_6$}{Unusual Valence of Ru and Prediction of Magnetism, Anomalous Hall Conductivity in a Newly Synthesized Double Perovskite Compound Ca_2CoRuO_6}}}

\author{Koushik Pradhan$^{\ast}$}
\email{Equal contribution first author}
 \affiliation{Department of Condensed Matter and Materials Physics,
S. N. Bose National Centre for Basic Sciences, Kolkata 700106, India}
\author{Soumya Ghorai$^{\ast}$}
\email{Equal contribution first author}
\affiliation{Department of Condensed Matter and Materials Physics,
S. N. Bose National Centre for Basic Sciences, Kolkata 700106, India}
\author{Prabuddha Sanyal}
\affiliation{Department of Physics, Maulana Abul Kalam Azad University of Technology, Nadia  741249, West Bengal, India.}
\author{Ryan Morrow}
 \affiliation{Institut f\"{u}r Festk\"{o}rperphysik, Leibniz IFW Dresden, D-01069 Dresden, Germany}
\author{Bernd B\"{u}chner}
 \affiliation{Institut f\"{u}r Festk\"{o}rperphysik, Leibniz IFW Dresden, D-01069 Dresden, Germany}
 \affiliation{Institut f\"{u}r Festk\"{o}rperphysik, TU Dresden, D-01062 Dresden, Germany}

\author{Thirupathaiah Setti}
\email{setti@bose.res.in}
  \author{Tanusri Saha Dasgupta}
 \email{t.sahadasgupta@gmail.com}
 \affiliation{Department of Condensed Matter and Materials Physics,
S. N. Bose National Centre for Basic Sciences, Kolkata 700106, India}
\date{\today}

\begin{abstract} With a goal to expand on the family of double perovskite compounds, hosting 3d transition metal and 4d or 5d transition metal, two new ordered double perovskite compounds,
Ca$_2$FeRuO$_6$ and Ca$_2$CoRuO$_6$ are synthesized following the prediction of a recent high throughput machine-learning study  [Phys. Rev. Materials 3, 084418]. Experimentally both compounds are found to stabilize in monoclinic symmetry, which is consistent with the high-throughput prediction for Ca$_2$FeRuO$_6$, but at odd for Ca$_2$CoRuO$_6$. Among
the two synthesized compounds, the properties of Ca$_2$CoRuO$_6$, investigated
employing the first principles technique \textcolor{black}{and model Hamiltonian calculation}, appear promising. 
The monoclinic structured Ca$_2$CoRuO$_6$ is found to stabilize unusual 6+ valence of Ru, and support a half-metallic ground state with uncompensated net moment.
As \textcolor{black}{predicted} by our first-principles study, the finite spin-orbit coupling 
at the Ru site contributes to the non-trivial topology of the band structure of monoclinic Ca$_2$CoRuO$_6$, resulting in a \textcolor{black}{moderately} large value of anomalous Hall conductivity. Our \textcolor{black}{theoretical predictions} should encourage further experimental investigation of this newly synthesized compound.

\end{abstract}

\maketitle

\section{Introduction}
Perovskite compounds of general formula ABO$_3$ have been the focus of the attention of materials physicists and chemists for ages due to diverse combinations of A (alkaline earth or rare-
earth metal) and B (transition metal or smaller lanthanide) cations that the structure 
can host. In recent years, perovskite-derived double perovskite compounds have been
discussed for various applications that include spintronics \cite{spintronics},
multiferroicity \cite{Oka2008}, magnetodielectric \cite{hena_prl}, magneto-optic devices \cite{hdas_apl,rvaidya_prb}, 
in renewable energy \cite{renewable}, to name a few. In the double perovskite structure,
exactly half of the B sites are substituted by another transition (TM )site, B$^{'}$
with rock-salt ordering between B and B$^{'}$, which doubles the unit cell and the 
formula to A$_2$BB$^{'}$O$_6$. The excitement on double perovskite compounds was initiated by the report of conducting ferromagnetism with magnetic transition temperature above room temperature of 410-450 K \cite{tokura_sfmo_nature} in Sr$_2$FeMoO$_6$. Subsequently, several conducting ferromagnetic double perovskites with high magnetic transition temperature have been
discovered \cite{hdas_prb_2011,sanyal_prb_2016,Serrate_2007,philipp,sahnoun2017magnetic}. Theoretically, many of them have been predicted to be half-metallic, opening the possibility of application as spintronics materials.

Among the double perovskite family, the compounds with 3d TM ion at the B site and 4d or 5d TM ion at
B$^{'}$ sites are particularly appealing due to the interplay of several 
different energy scales involving bandwidth, Coulomb correlation, and spin-orbit coupling (SOC) of the 3d and 4d/5d orbitals. 
While the 3d TM atoms at the B site act as hosts for high energy scale for 
magnetism, the 4d or 5d TM atoms at B$^{'}$ site feature SOC.
Separation of the ions in B and B$^{'}$ sites that host magnetism from the ions
which host SOC, avoids the disadvantageous issues related
to the counter effect of correlation and SOC at the same site,
as in the case of single perovskite compounds. This offers a unique platform
to give rise to fascinating topological properties with real-life application possibilities for example in low-dissipation devices \cite{coey2001materials}. Thus, there is an urgent need to expand on this family of compounds.

In recent years, about 30 new double perovskites with 3d and 4d or 5d TMs at B and B$^{'}$ sites have been predicted employing the power of machine learning and high throughput calculation \cite{anita_prm}. Based on this, in this study, we report the successful synthesis of two Ru-based double perovskites, Ca$_2$CoRuO$_6$ (CCRO) and Ca$_2$FeRuO$_6$ (CFRO). It is worth mentioning at this point, that synthesis of CFRO compound has been previously reported in literature \cite{Naveen2018}. However, the space group
of the synthesized compound was found to be orthorhombic, suggesting the absence of ordering
between Fe and Ru. A spin-glass-like transition around 87 K was reported possibly due to the disorder arising from random occupancy of iron and ruthenium atoms at the B site.
On the other hand, in the presently synthesized compounds, as elaborated below, ordering is achieved with some antisite disorder.
Following the successful synthesis of the two compounds, we take a critical look
at the physical properties of the two compounds, \textcolor{black}{primarily employing theoretical calculations}. In the process, we discover an 
unusual valence for octahedrally coordinated Ru ion in CCRO.
\textcolor{black}{Based on this, we predict}
stabilization of a net-moment half-metallic state, and \textcolor{black}{a reasonably high} value of intrinsic anomalous Hall
conductivity (AHC), arising from topological band crossings in the 
proximity of the Fermi energy. The AHC in CCRO is found to be comparable to others predicted 
recently \textcolor{black}{for 3d-4d/5d double perovskite\cite{samanta_npj_2023large}, thereby expanding on this material class as promising topological compounds}. CFRO on the other hand, turns out to be a trivial ferrimagnetic insulator. We also uncover the mechanism behind the stabilization of the unusual valence of
Ru in CCRO. It is interesting to note that while the predicted crystal symmetry 
of CFRO in high throughput calculation agrees with the experimentally determined
one, they differ for CCRO. The more precise calculation, reported below, supports the experimentally determined crystal symmetry.
\textcolor{black}{The experimentally determined crystal structure
turns out to be crucial for the stabilization of the unusual valence of Ru. This is found to greatly influence the calculated physical properties.} This stresses the
importance of the feedback between experiments and calculations in 
discovering new compounds with exciting properties.

\section{Method}

\subsection{Experimental}
Polycrystalline powders of Ca$_{2}$FeRuO$_{6}$ and Ca$_{2}$CoRuO$_{6}$ were synthesized via the conventional solid-state reaction method. High-purity (\(\geq\)99.9\%) precursors of CaCO$_{3}$ (Sigma Aldrich), Fe$_{2}$O$_{3}$ (Thermo Scientific), Co$_{3}$O$_{4}$ (Alfa Aesar), and RuO$_{2}$ (Alfa Aesar) were used in stoichiometric proportions. The precursors were thoroughly mixed using an agate mortar and pestle for approximately 2 hours to ensure homogeneity. The mixture was then loaded into an alumina crucible and subjected to an initial heat treatment at 600$^\circ$C for 24 hours to minimize Ru evaporation. Subsequent heat treatments were performed at higher temperatures, with intermittent grinding, until the desired phase purity was achieved. High-pressure annealing was conducted using an MTI OTF-1200X-HP-55 split-tube furnace. To synthesize Ca$_{2}$FeRuO$_{6}$ two routes were taken. In the first attempt, the sample was annealed at 1100$^\circ$C in air, followed by furnace cooling. In the second, the sample was annealed at 900$^\circ$C under continuous O$_{2}$ flow for 36 hours, followed by controlled cooling at a rate of 50$^\circ$C/h. For Ca$_{2}$CoRuO$_{6}$, the mixture was pre-reacted at 900$^\circ$C under continuous O$_{2}$ flow for 24 hours, with a cooling rate of 50$^\circ$C/h. The final annealing step was carried out at 1000$^\circ$C under a constant O$_{2}$ pressure of 5MPa for 24 hours, followed by cooling at 100$^\circ$C/h.

Phase purity and structural analysis were performed using powder X-ray diffraction (XRD). The XRD measurements were conducted in transmission geometry on a STOE STADI diffractometer with Co K$_{\alpha}$ ($\sim$1.79\AA) radiation, equipped with a germanium monochromator and a DECTRIS MYTHEN 1K detector. Additionally, the orthorhombic Ca$_{2}$FeRuO$_{6}$ phase was characterized using a PANalytical X-PERT PRO diffractometer with Cu K$_\alpha$ ($\sim$1.54\AA) radiation. The obtained XRD patterns were analyzed via Rietveld refinement using \textsc{Jana-2020}~\cite{Petricek2023}. \textcolor{black}{Three-dimensional visualization of the refined crystal structure, generated using VESTA software~\cite{Momma2008}. Further details on synthesis can be found
in the supplementary information (SI) section~\ref{sec:SI}.
The direct current (DC) magnetization measurements of Ca$_2$CoRuO$_6$ were carried out using a Quantum Design SQUID MPMS. For the experiment, the polycrystalline sample was enclosed in size 4 gel capsule, which was then secured inside a plastic straw and mounted to a standard sample stick for insertion into the device. Magnetization as a function of temperature was recorded under applied magnetic fields of 100 Oe and 10000 Oe, using both zero-field-cooled (ZFC) and field-cooled (FC) measurement protocols. The measurements were conducted over the temperature range of 2-350 K.}

\subsection{Computational}

The first-principles density functional theory (DFT) calculations were carried out in the plane-wave basis with projector augmented-wave potentials \cite{paw}, as implemented in the Vienna Ab initio Simulation Package ( VASP ) \cite{vasp_prb_93,vasp_prb_96}. The exchange-correlation functional was approximated within the generalized gradient approximation (GGA) of Perdew-Burke-Ernzerhof \cite{pbe_prl_96}. The electron-electron correlation at the transition-metal sites beyond the level of GGA was
taken into account through a supplemented onsite Coulomb repulsion (U) and Hund exchange J$_H$ correction as implemented in Liechtenstein's multiorbital GGA+U formalism \cite{lichtenstein_prb_95}. Within the GGA+U formulation of Liechtenstein,\cite{lichtenstein_prb_95}  U and J$_H$ are two parameters of the theory for which choices need to be made. Following the 
literature \cite{hubbardu_prb_22,hubbardu_prb_94}, we considered U=6 eV for the Fe/Co site and U=1 eV for the Ru site. The Hund coupling J$_H$ was chosen to be 0.8 eV for both Fe/Co and 0.5 eV for Ru sites. The U parameter was varied over 1-2 eV, and
the qualitative trend was found to remain unchanged. A plane-wave energy cutoff of 600 eV and Brillouin zone sampling with $6\times6\times4$ Monkhorst-Pack grids were sufficient for the convergence of energies and forces. For structural relaxations, ions were allowed to move until the atomic forces became less than 0.0001 $eV/\AA$.

To extract the few-band tight-binding Hamiltonian from the full DFT calculation, which serves as input for the model calculation, we performed Nth Order Muffin-Tin Orbital (NMTO)\cite{nmto_prb_2000} downfolding calculations. The NMTO method, which is not yet available in its self-consistent form, relies on the self-consistent potential parameters derived from Linear Muffin-Tin Orbital (LMTO)\cite{lmto_prl_84} calculations. The results were cross-verified by comparing the density of states (DOS) and band structures obtained from plane wave and LMTO calculations.

The topological properties were calculated using the WANNIER90 code\cite{wannier90}. To do this, we first computed maximally localized Wannier functions (MLWF) to derive a tight-binding model from ab initio DFT calculations in the Co-d-Ru-t$_{2g}$ basis. With the help of WannierTools\cite{wanntools}, the topological nature of the energy bands near the Fermi level was analyzed by calculating the Chern number and anomalous Hall conductivity.

Berry curvature in the clean limit was obtained by using the Kubo formalism,\cite{Kubo} expressed in the following
equation,
\begin{equation}
\Omega_{n}^{z}(k)=-2Im\sum\limits_{m\not=n}\dfrac{<n|\hat{v_x}|m><m|\hat{v_y}|n>}{({\epsilon_{m}-\epsilon_{n}})^2}
\label{eqn:berry}
\end{equation}
where, $\epsilon_m$ and $\ket{m}$ are the m-th energy eigenvalue and eigenvector of the Hamiltonian, $\Hat{v_x}$ and $\hat{v_y}$ are the velocity operators along x and y respectively.

Chern number was estimated by integration of Berry curvature over the first Brillouin Zone,\textcolor{black}{\cite{Kubo}} as given below.
\begin{equation}
    C_{N}=\dfrac{1}{2\pi}\sum\limits_{n}\int_{BZ} \textcolor{black}{\Omega_{n}^{z}(k)} d^2k
\label{eqn:chern}
\end{equation}
where \textcolor{black}{$\Omega_{n}^{z}(k)$} is the berry curvature of the $n$-th band at a specified k point.

Anomalous hall conductivity, arising from the Berry Curvature, was computed employing the 
following equation,\textcolor{black}{\cite{Kubo}}
\begin{equation}
\sigma_{xy}=-\dfrac{e^2}{\hbar}\sum\limits_{n}\int_{BZ}\dfrac{d^3k}{(2\pi)^3}\Omega_{n}^{z}(k)f_{n}
\label{eqn:ahc}
\end{equation}
where $f_n$ is the Fermi-Dirac distribution function and $\hbar$ is the reduced Planck's constant.

\section{Physical Properties}
\subsection{Experimental Structures}

\begin{figure}[ht]
        \centering
	\includegraphics[width=0.95\linewidth, clip=true]{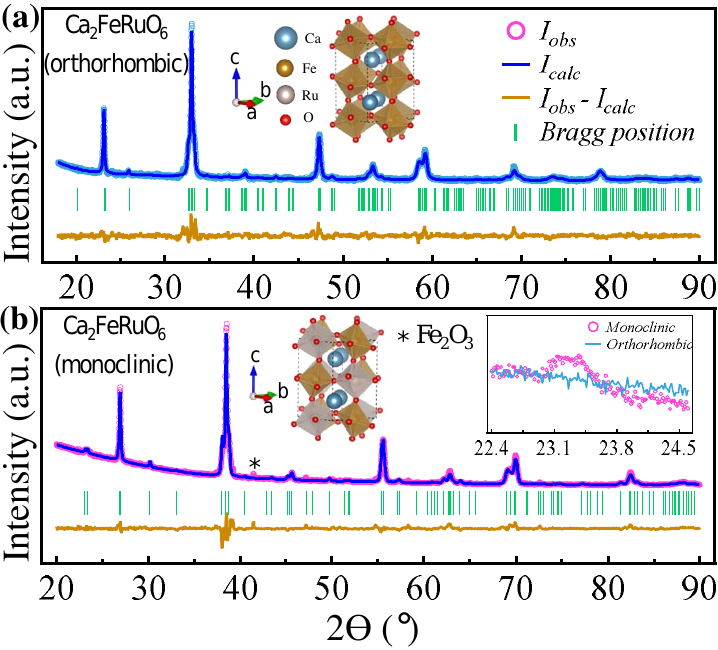}
	\caption{\textcolor{black}{Rietveld refinement profile of the powder X-ray diffraction pattern for (a) Ca$_{2}$FeRuO$_{6}$ (orthorhombic phase), and (b) Ca$_{2}$FeRuO$_{6}$ (monoclinic phase). Inset of (b) shows the difference of R-point reflection in the XRD pattern for monoclinic and orthorhombic phases of Ca$_{2}$FeRuO$_{6}$.}}
	\label{fig-exp1}
\end{figure}

\begin{figure}[ht]
        \centering
	\includegraphics[width=0.95\linewidth, clip=true]{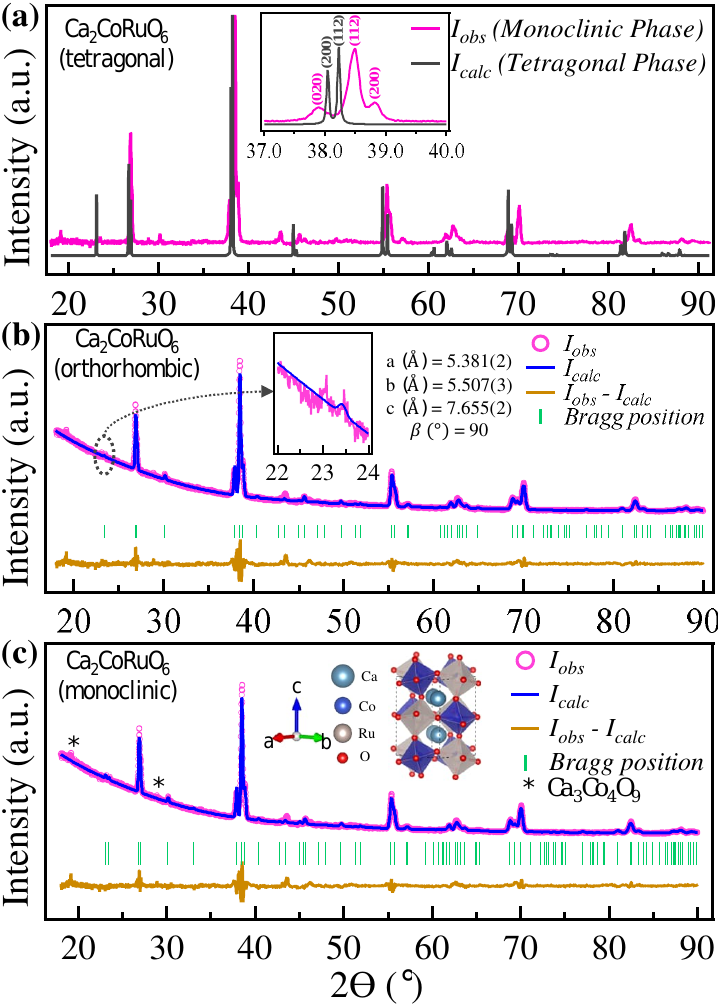}
	\caption{\textcolor{black}{(a) Shows the comparison between monoclinic and tetragonal phase of Ca$_{2}$CoRuO$_{6}$. Rietveld refinement profile of the powder X-ray diffraction pattern for (b) Ca$_{2}$CoRuO$_{6}$ (orthorhombic phase), and (c) Ca$_{2}$CoRuO$_{6}$ (monoclinic phase). Inset of (b) shows the difference between experimental XRD pattern and Rietveld refinement profile in the R-point region for orthorhombic phase of Ca$_{2}$CoRuO$_{6}$.}}
	\label{fig-exp2}
\end{figure}

Room-temperature X-ray powder diffraction (XRD) was employed to determine the crystal structures of Ca$_{2}$FeRuO$_{6}$ and Ca$_{2}$CoRuO$_{6}$. \textcolor{black}{Figs.~\ref{fig-exp1}, ~\ref{fig-exp2}(b), and  ~\ref{fig-exp2}(c)} present the XRD patterns overlapped with their respective Rietveld refinements. The most common arrangement of B and B$^{'}$ cations in ordered double perovskite
is the rock salt arrangement, in which BO$_6$ and B$^{'}$O$_6$ octahedra are corner-shared and alternately arranged in all three crystallographic directions.
In the rock salt structure, the ideal cubic ordered double perovskite (DP) has a space group symmetry of \( \text{Fm-3m} \). Group theoretical studies \cite{Woodward:br0052} show that octahedral tilting, which depends on the
Goldschmidt's tolerance factor $t$ \cite{Goldschmidt1926}, $ t = \frac{r_A + r_O}{\sqrt{2}(\frac{r_B + r_{B'}}{2} + r_O)} $ ($r_A$, $r_B$, $r_{B'}$ and r$_O$ represent the ionic radii of the A cation, B cation, B$^{'}$ cation and O$^{2-}$ anion)
can reduce the symmetry to specific space groups. Literature \cite{VASALA20151} shows that the tolerance factor of most DP compounds falls within the range \( t \sim 0.93 - 1.01 \). Only a few compounds are found with \( t > 1.01 \), where the A-site cation is too large for the perovskite structure, leading to the formation of various non-perovskite hexagonal structures. The synthesized DPs so far are 
reported in space groups of cubic \( \text{Fm-3m} \), rhombohedral \( \text{R-3} \), tetragonal \( \text{I4/m} \), tetragonal \( \text{I4/mmm} \), monoclinic \( \text{P2$_1$/n} \), monoclinic \( \text{I2/m} \), and rare cases such as tetragonal \( \text{P4/mnc} \) and monoclinic \( \text{I-1} \). The tolerance factor
of CCRO and CFRO is expected to be in the range 0.96-0.97, for which monoclinic
\( \text{P2$_1$/n} \) as well as tetragonal \( \text{I4/m} \) symmetries are possible. 

\textcolor{black}{Fig.~\ref{fig-exp1}(a)} shows the Rietveld refinement profile of Ca$_{2}$FeRuO$_{6}$ assuming an orthorhombic space group of  \text{Pbnm} (No. 62) with the refined structural parameters \(\chi^2 = 1.53, R_p = 2.21, R_{wp} = 2.97, R^2 = 4.94\) and the lattice constants \text{a} = 5.384(5)~\AA, \text{b} = 5.485(7)~\AA, and \text{c} = 7.669(1)~\AA. These lattice parameters are in agreement with a previous report on the same system~\cite{Naveen2018}. The orthorhombic distortion in these systems arises from the octahedral rotations about the $[011]_{\text{cubic}}$ and $[100]_{\text{cubic}}$ axes, which optimizes only eight out of the initial twelve equal A-O bond lengths, leading to the absence of \( 0kl: k = 2n + 1 \) reflections~\cite{Anderson1993}. In this phase, Ca ions occupy the 4c Wyckoff positions (\textit{x},\textit{y},$\frac{1}{4}$), while Fe and Ru ions are randomly distributed at the 4b Wyckoff positions ($\frac{1}{2}$,0,0).

However, the recent high throughput calculations suggested Ca$_{2}$FeRuO$_{6}$ to be in the monoclinic phase with a space group of \text{P2$_{1}$/n} (No. 14)~\cite{anita_prm}. 
We note that while the orthorhombic symmetry indicates a disordered distribution
of B-site cations, Fe and Ru, the monoclinic symmetry indicates a rock-salt ordering of B-site Fe and B$^{'}$-site Ru ions. The charge difference \(\Delta Z_B = 2\) between Fe$^{3+}$ and Ru$^{5+}$, coupled with an ionic radius mismatch \(\Delta r_B \approx 2\)~\AA, enhances the likelihood of B-site cation ordering~\cite{VASALA20151}. 
Motivated by this, we optimized the synthesis conditions by lowering
the annealing temperature from 1100\textdegree C to 900\textdegree C with a hope to stabilize Ca$_{2}$FeRuO$_{6}$ in the monoclinic phase. \textcolor{black}{Fig.~\ref{fig-exp1}(b)} shows the XRD pattern of Ca$_{2}$FeRuO$_{6}$ overlapped with the Rietveld refinement, assuming the monoclinic \text{P2$_{1}$/n} (No. 14) symmetry. The fit is found
to be very good with refined parameters \(\chi^2 = 1.56, R_p = 2.22, R_{wp} = 4.02, R^2 = 7.43\) and lattice constants \textit{a} = 5.386(5)~\AA, \textit{b} = 5.484(8)~\AA, \textit{c} = 7.664(3)~\AA, and $\beta$ = 90.016\textdegree. 
The monoclinic distortion, similar to the orthorhombic case, arises due to octahedral rotations about the $[011]_{\text{cubic}}$ and $[100]_{\text{cubic}}$ axes but with the presence of \( 0kl: k = 2n + 1 \) reflections~\cite{Anderson1993}. In the monoclinic phase, the Ca ions occupy the 4e Wyckoff positions (\textit{x},\textit{y},\textit{z}), while the Fe and Ru ions occupy the 2c (0,$\frac{1}{2}$,0) and 2d ($\frac{1}{2}$,0,0) Wyckoff positions, respectively, forming a checkerboard-type ordering of FeO$_6$/RuO$_6$ octahedra in the face-centred cubic lattice.
The relationship with cubic lattice parameter $a_p$ (\(\sqrt{2}a_p \times \sqrt{2}a_p \times 2a_p\)) remains identical for orthorhombic and monoclinic structures, leading to nearly identical Bragg peak positions. This makes it challenging to distinguish one symmetry from the other. Nevertheless, one can distinguish these two phases by closely observing the R-point reflections in the monoclinic symmetry and its
absence in the orthorhombic symmetry. P. W. Barnes \textit{et al.}~\cite{Barnes2006} categorized Bragg's reflections in the double perovskites into four groups, among which the R-point reflections indexed by all-odd Miller indices are possible indications of B-site cation ordering. 
We observe that synthesis at a lower temperature (900\textdegree C) results in significant intensity at the R-point position [see inset of \textcolor{black}{Fig.~\ref{fig-exp1}(b)}], which diminishes with synthesis at high temperature (1100\textdegree C). This establishes
the absence of cation ordering in an earlier report was due to the high temperature synthesis condition of 1280\textdegree C~\cite{Naveen2018} and thus highlights the crucial role of synthesis temperature in cation ordering.

\textcolor{black}{The XRD structural characterization of Ca$_{2}$CoRuO$_{6}$ is shown in Fig.~\ref{fig-exp2}. Interestingly, while theoretical predictions suggested a tetragonal structure for Ca$_{2}$CoRuO$_{6}$~\cite{anita_prm}, experimentally this system is found to crystallize in different structure. Fig.~\ref{fig-exp2}(a) shows experimentally obtained  XRD pattern of Ca$_{2}$CoRuO$_{6}$ overlapped with the theoretically predicted tetragonal structure of space group \text{I4/m} (simulated using VESTA with the help of predicted structure). The Bragg plane's peak positions of the experimental data (indexed to a monoclinic phase) are different from the assumed tetragonal phase. The simulated XRD pattern of the tetragonal phase does not exhibit certain reflections that are present in the experimental XRD pattern. This difference is highlighted in the inset of Fig.~\ref{fig-exp2}(a).}

\textcolor{black}{At a first glance, the XRD pattern of Ca$_{2}$CoRuO$_{6}$ might appear to be a disordered structure rather than a rock-salt ordered double perovskite structure, due to the weak R-point reflections. To clarify this further we have performed the Rietveld refinement as shown in Fig.~\ref{fig-exp2}(b), assuming an orthorhombic \text{Pbnm} (No. 62) space group (see section ~\ref{sec:suppsec2}). From the inset of Fig.~\ref{fig-exp2}(b), it is evident that the orthorhombic phase has only one peak at 23.43 corresponding to the (101) plane, while the experimental XRD pattern has two peaks at the R-point reflection region. Finally, in Fig.~\ref{fig-exp2}(c), we show the XRD pattern of Ca$_{2}$CoRuO$_{6}$ overlapped with the Rietveld refinement, confirming the monoclinic phase with a space group of \text{P2$_{1}$/n} (No. 14) with the refined parameters \(\chi^2 = 1.54, R_p = 1.88, R_{wp} = 3.21, R^2 = 5.14\) and lattice constants \text{a} = 5.381(3)~\AA, \text{b} = 5.506(8)~\AA, \text{c} = 7.654(5)~\AA, and $\beta$ = 90.018\textdegree. This demonstrates that our synthesized sample Ca$_{2}$CoRuO$_{6}$ indeed has the rock-salt-type ordering of Co and Ru with a monoclinic space group of \text{P2$_{1}$/n}.} 

\textcolor{black}{We would like to stress here that significant effort was made to synthesize phase-pure samples by considering three different synthesis protocols, as elaborated in sec~\ref{sec:suppsec1}. However, all different attempts, resulted in samples containing a fraction of impurity phase, together with Ca$_2$CoRuO$_6$. In particular, an impurity phase of the misfit cobaltate compound Ca$_3$Co$_4$O$_9$ is identified, with its diffraction peaks marked by asterisks in Fig.~\ref{fig-exp2}(c). We further quantified the antisite disorder in the synthesized  Ca$_{2}$FeRuO$_{6}$ and Ca$_{2}$CoRuO$_{6}$ systems to be approximately \(25\% \pm 5\%\) assuming the different site occupancies of B/B$'$ cation. See Fig~\ref{fig4}  for the refinements of Ca$_{2}$CoRuO$_{6}$ considering different site occupancies of Co and Ru ions in the 2c and 2d Wyckoff positions within the monoclinic \text{P2$_{1}$/n} structure.} 

\subsection{Theoretically Optimized Crystal Structures}

\begin{figure*}
\centering
    \includegraphics[scale=0.5]{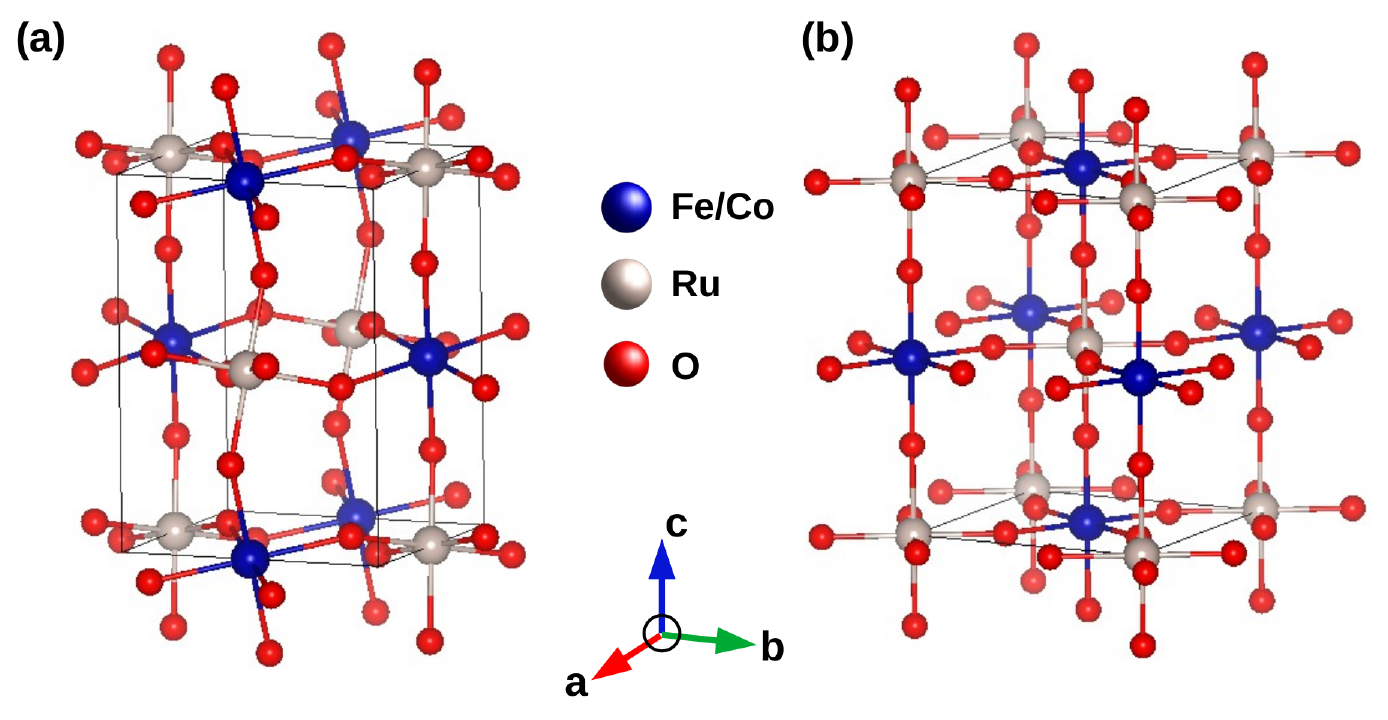}    
    \caption{The crystal structures of  monoclinic and tetragonal symmetries. Ca atoms are omitted due for the clarity of the picture. In both structures, the octahedra formed by Fe or Co atoms at B sites of the double perovskite structure, corner share with octahedra formed by Ru atoms at B$^{'}$ sites, with BO$_6$ and B$^{'}$O$_6$ octahedra alternating in all three directions. The octahedral tilt pattern in monoclinic
    structure is a$^-$a$^-$c$^+$, while that in tetragonal structure is  a$^0$a$^0$c$^-$.}
        \label{fig1:struct}
\end{figure*}

The high throughput calculation\cite{anita_prm} predicted the monoclinic 
symmetry for CFRO which is in agreement with the experiment, and tetragonal symmetry for CCRO which is in contrast with the experiment. The octahedral tilt
pattern is a$^-$a$^-$c$^+$ and a$^0$a$^0$c$^-$ for the monoclinic space group (No. 14) and
tetragonal space group (No. 87) respectively, as per Glazer notation (cf Fig.\ref{fig1:struct}).
The lowering of symmetry from tetragonal to monoclinic symmetry removes the C$_4$ (4-fold rotation axis), $m$ (mirror plane perpendicular to the 4-fold axis)
 symmetry elements.  
 
To clarify this issue, we performed DFT optimization assuming 
both tetragonal and monoclinic structures of CCRO and CFRO. In the optimization
process, the volume and the ionic positions were relaxed.  Total energy associated with the optimized
structures reveals that CFRO favours the monoclinic structure, with an energy 
0.4 eV/f.u lower than the tetragonal structure, in agreement with both experimental measurement and high throughput prediction. CCRO, on the other hand, is found to stabilize in the monoclinic symmetry, with an energy of 0.7 eV/f.u lower than that of the tetragonal structure, confirming the experimental observation.

The optimized structural parameters for CCRO and CFRO in their respective space groups are summarized in Table.\ref{tab:struct}.

\begin{table*}
\begin{center}
\begin{tabular}{|c|c c c|c c c|}
\hline
 & \multicolumn{3}{c|}{CCRO} & \multicolumn{3}{c|}{CFRO} \\
\hline
 & Monoclinic (theory) & Monoclinic (expt) & Tetragonal (theory) & Monoclinic (theory) & Monoclinic (expt) & Tetragonal\\
 & P2$_1$/c & P2$_1$/c & I4/m & P2$_1$/c  & P2$_1$/c & I4/m  \\
a($\AA$) & 5.429 & 5.381 & 5.495 & 5.408 & 5.386 & 5.432 \\
b($\AA$) & 5.627 & 5.507 & 5.495 & 5.560 & 5.485 & 5.432 \\
c($\AA$) & 7.714 & 7.654 & 7.702 & 7.710 & 7.664 & 7.897 \\
V($\AA^3$) & 235.636 & 226.831 & 232.555 & 231.843 & 226.433 & 233.036 \\
$\alpha$($^0$) & 90 & 90 & 90 & 90 & 90 & 90 \\
$\beta$($^0$) & 89.858 & 90.018 & 90 & 90.031 & 90.016 & 90\\
$\gamma$($^0$) & 90 & 90 & 90 & 90 & 90 & 90 \\
Co-O($\AA$) & 2.072($\times2$) & 1.932($\times2$)  & 1.886($\times2$) & 2.012($\times2$) & 1.939($\times2)$  & 1.989($\times2$)\\
& 2.101($\times2$)& 2.020($\times2$) & 2.059($\times4$) & 2.027($\times2$) & 2.083($\times2$)  & 2.004($\times4$)\\
& 2.125($\times2$)& 2.126($\times2$)  & & 2.032($\times2$) & 2.126($\times2$)  &\\
Ru-O($\AA$) & 1.943($\times2$) & 1.890($\times2$)  & 1.962($\times2$) & 1.946($\times2$) & 1.953($\times2$)  & 1.959($\times2$)\\
& 1.948($\times2$)& 1.931($\times2$)  & 1.965($\times4$) & 1.982($\times2$) & 1.974($\times2$)  & 1.969($\times4$)\\
& 1.955($\times2$)& 1.988($\times2$)  & & 1.983($\times2$) & 1.808($\times2$)  & \\
$\langle$B-O-Ru($^0$)& 149.059 & 143.164 & 150.109($\times2$) & 150.014 & 162.103 & 150.398($\times2$) \\
            & 147.288 & 158.317 & 180 & 150.698 & 140.856 & 180 \\
            & 147.649 & 156.234 & & 150.406 & 156.558 & \\
\hline
\end{tabular}
\end{center}
\caption{The structural details of CCRO and CFRO in monoclinic and tetragonal 
spacegroups. Listed are the structural parameters of the theoretically optimized
structures along with refined experimentally structures, which are found to be monoclinic for both CCRO and CFRO.}
\label{tab:struct}
\end{table*}

 In the case of CCRO, the Co-O bond lengths show a significant difference between the tetragonal and monoclinic phases, increasing by approximately 5\% in the monoclinic phase. The Ru-O bond lengths, though, remain unchanged up to the first decimal place. This variation in the Co-O bond lengths could potentially change the predicted electronic and magnetic properties of ground-state monoclinic structured CCRO, as found below. On the other hand, for CFRO, both the Fe-O and Ru-O bond lengths remain nearly unchanged between the two space groups.

\subsection{Electronic Structure and Magnetic Ground State}

\begin{figure*}
\centering
    \includegraphics[scale=0.6]{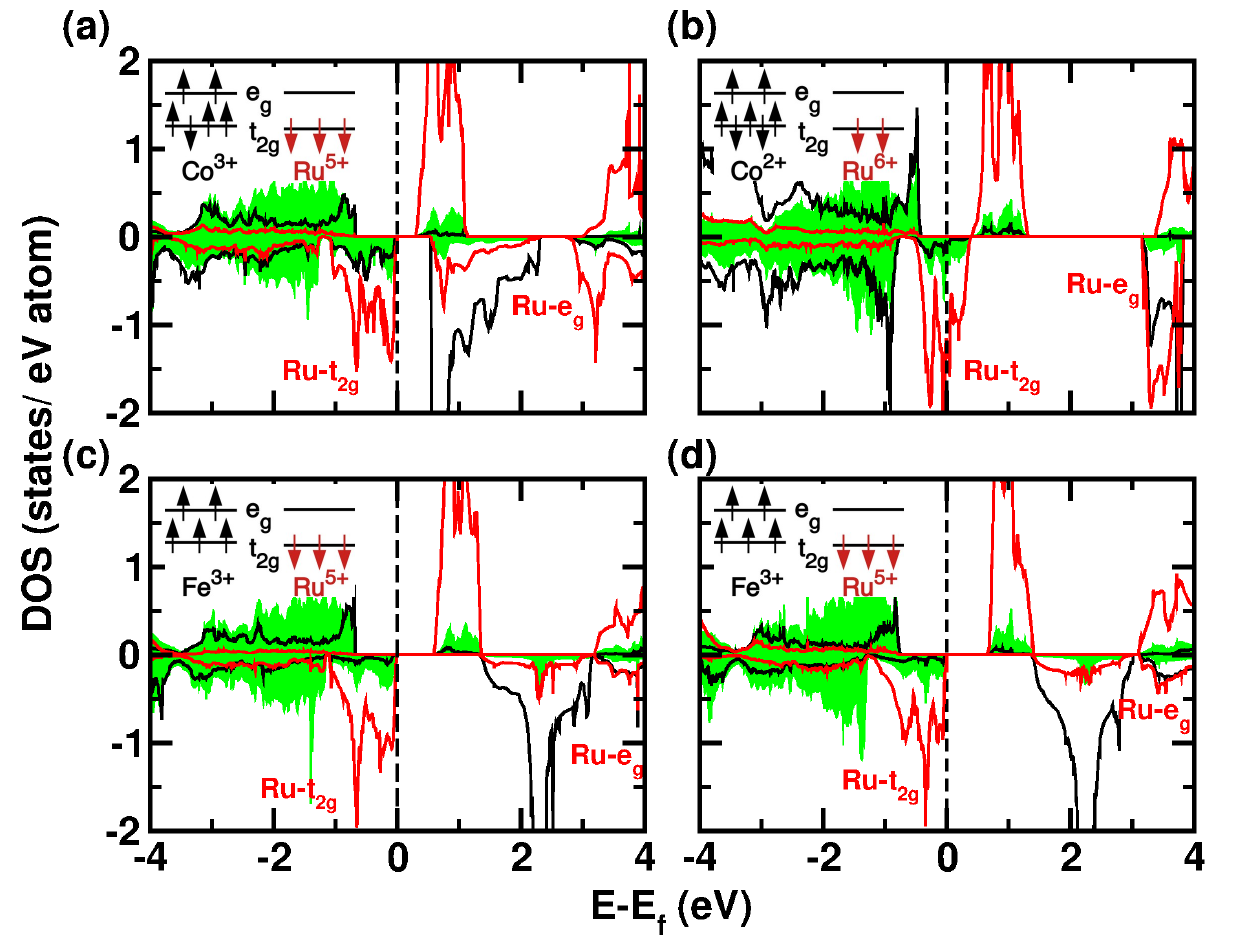}    
    \caption{The spin polarized GGA+U density of states projected on Co/Fe-d (black solid line), Ru-d (red solid line) and O-p (green shaded) states. Zero of the energy is set at Fermi level. Panels (a), (b) show the results for CCRO in tetragonal and monoclinic phases respectively. Panels (c), (d) show the same for CFRO. Insets show the electron configurations of Co/Fe and Ru.}
        \label{fig2:dos}
\end{figure*}

To understand the influence of crystal symmetry on the electronic structure,
we first calculated the magnetic moments and density of the states (DOS) of CCRO and CFRO, assuming both the tetragonal and monoclinic phases. The density of states
projected to Co/Fe-d, Ru-d and O-p states are shown in Fig.\ref{fig2:dos} and
the atom projected magnetic moments are listed in Table \ref{tab:moment}.
\textcolor{black}{The calculations have been carried out considering both
experimental crystal structure as well as theoretically optimized structure. The results are found to be similar.}

Comparing the DOS of CCRO and CFRO in the two assumed crystal symmetries,
a marked difference is noticed for CCRO while features are found 
similar for CFRO.
While CFRO both in tetragonal and monoclinic symmetry led to an insulating solution
with occupied Fe-d states in the up (majority) spin channel and occupied crystal field split Ru t$_{2g}$ states in the down (minority) spin channel, the scenario
is found to be remarkably different between tetragonal CCRO and monoclinic CCRO.
The density of states of tetragonal CCRO is insulating with occupied Fe-d states in the up (majority) spin channel and occupied crystal field split Ru t$_{2g}$ states along with some of Co t$_{2g}$ states in the down (minority) spin channel. The density of states of monoclinic CCRO, on the other hand, is found to be half-metallic with a gap of $\sim$ 1 eV in the majority spin between occupied Co-d states and unoccupied Ru t$_{2g}$ states, and the Ru t$_{2g}$ states admixed
with Co-d and O-p crossing the Fermi level in the minority spin channel. 
\begin{table*}
\begin{center}
\begin{tabular}{ |c |c c c|c c c|}
\hline
 & \multicolumn{3}{c|}{CCRO} & \multicolumn{3}{c|}{CFRO} \\
\hline
 & Monoclinic (theory) & \textcolor{black}{Monoclinic (expt)} & Tetragonal (theory) & Monoclinic (theory) &  \textcolor{black}{Monoclinic (expt)} & Tetragonal\\
 & P2$_1$/c  & \textcolor{black}{P2$_1$/c} & I4/m & P2$_1$/c & \textcolor{black}{P2$_1$/c} & I4/m  \\
$\mu_{B}$ & 2.73 &\textcolor{black}{2.83} &3.1 & 4.26 &\textcolor{black}{4.22} &4.29 \\
          & (2.67) & \textcolor{black}{(2.57)}&(3.09) & (4.15)&\textcolor{black}{(4.03)}  & (4.17) \\ 
$\mu_{Ru}$ & -1.23 &\textcolor{black}{-1.43} &-1.71 & -1.76 &\textcolor{black}{-1.65} &-1.76 \\
           & (-1.12)&\textcolor{black}{(-1.23)} &(-1.61) & (-1.65)&\textcolor{black}{(-1.44)} &(-1.65) \\
$\mu_O$ & -0.07 &\textcolor{black}{-0.09} &-0.06 & -0.05 &\textcolor{black}{-0.14} &-0.06 \\
          &(-0.06) &\textcolor{black}{(-0.07)} &(-0.07) & (-0.05)&\textcolor{black}{(-0.13)}  & (-0.05) \\ 
$\mu_{total}$ & 1 &\textcolor{black}{1} &1 & 2&\textcolor{black}{2} & 2 \\
          &  (1) &\textcolor{black}{(1)} &(1) & (2)&\textcolor{black}{(2)}  & (2) \\ 
\hline
\end{tabular}
\end{center}
\caption{\textcolor{black}{Magnetic moments estimated by DFT calculation with the plane-wave basis using both experimental and theoretically optimized crystal structures. The values in the parentheses are obtained using linearized augmented plane-wave (LAPW) basis. All the values are in the unit of $\mu_B$.}}
\label{tab:moment}
\end{table*}

This is also reflected in the calculated magnetic moments. The magnetic
moments at the Fe/Co site and Ru site are found to be oppositely oriented
in all cases, with a net magnetic moment of 1 $\mu_B$ for CCRO and of 2 $\mu_B$
for CFRO. For CFRO, the magnetic moments at Fe and Ru sites are found to be 
similar between monoclinic and tetragonal phases with about m$_{Fe}$ $\simeq$ \textcolor{black}{4.2-4.3} 
$\mu_B$, and m$_{Ru}$ $\simeq$ \textcolor{black}{- 1.7-1.8} $\mu_B$ with the rest of the moments at O sites
and interstitial. This leads to nominal valence of Fe$^{3+}$ with $d^5$ occupancy
and Ru$^{5+}$ with $d^3$ occupancy, as shown in insets of Fig-\ref{fig2:dos}(c)
and (d). The Fe and Ru spins of  S = 5/2 and S = 3/2 are aligned antiparallel due to antiferromagnetic superexchange (SE) between the half-filled Fe $d$ and Ru t$_{2g}$ states. This results in a ferrimagnetic configuration with a net moment of 2 $\mu_B$ per unit cell. The half-filled Fe $d$ and Ru t$_{2g}$ states contribute to the insulating behaviour of the system. The situation changes
drastically for CCRO. The magnetic moments at Co as well as that at Ru sites show
a significant reduction in the monoclinic phase compared to the tetragonal phase, although
the net moment is 1 $\mu_B$ in both cases. Assuming the nominal valence
of Ca$^{2+}$ and O$^{2-}$, the nominal valence of the Co-Ru combination can be
either 3+/5+ or 2+/6+. For tetragonal CCRO, the nominal 3+/5+ valences of 
B/B$^{'}$ is stabilized as in CFRO. The calculated magnetic moment of 3.1 $\mu_B$ suggests the high-spin configuration of Co$^{3+}$ d$^6$. The magnetic moment of 1.71 $\mu_B$ at the Ru site, which is oppositely oriented to the Co moment, is similar
to that of m$_{Ru}$ of CFRO, suggestive of low-spin Ru$^{5+}$ valence. The tetragonal symmetry of the crystal structure, driven by the Jahn-Teller effect of high-spin Co$^{3+}$ (d$^6$) in octahedral coordination, leads to significant differences between the in-plane and out-of-plane metal-oxygen bond lengths in CoO$_6$ (cf Table.\ref{tab:struct}). This distortion opens a gap in the 
down-spin channel of the Co t$_{2g}$ manifold, while in the up-spin channel, a gap is formed between filled Co $d$ and empty Ru t$_{2g}$ states, thereby inducing insulating behaviour in the system (cf Fig.\ref{fig2:dos}(a)). The change
in spacegroup from tetragonal to monoclinic symmetry in CCRO brings a major 
change. The Co magnetic moment changes from 3.1 $\mu_B$ to \textcolor{black}{2.7-2.8} $\mu_B$, resulting 
in higher d shell occupancy for Co, suggestive of high-spin d$^{7}$ occupancy
as opposed to high spin d$^{6}$. This makes Ru t$_{2g}$ states less than half-filled with a moment of \textcolor{black}{1.2-1.4} $\mu_B$ as opposed to 1.7 $\mu_B$ in the tetragonal phase.
The nominal valences of Co/Ru thus change from 3+/5+ in tetragonal CCRO
to 2+/6+ in monoclinic CCRO, as shown in insets of Fig. $\ref{fig2:dos}$(a),(b).
To the best of our knowledge, this is perhaps the first example of stabilization
of unusual valence of Ru in 6+ in octahedral coordination, though Ru$^{6+}$ has been reported in tetrahedral coordination in other types of compounds \cite{k2ruo4}. \textcolor{black}{To confirm the 6+ valence of Ru, we further carried out bond valence sum (BVS) analysis using the refined Co-O and Ru-O bond lengths obtained from Rietveld refinement of the X-ray diffraction data (cf Table I). For details see \ref{sec:suppsec3}. (2+,6+) description is found to fit the BVS much better compared to (3+,5+) description, supporting the theoretical findings. The deviation from (3+,5+) is found to be about twice larger than that of (2+,6+).}

The curious case of stabilization of Ru 6+ valence in CCRO may be \textcolor{black}{tentatively} rationalized in the following manner, by comparing the energy stability of Co 2+, Ru 6+ valence
vis-a-vis that of Cu 3+, Ru 5+ valence states in CCRO. As shown in the insets of
Fig.$\ref{fig2:dos}$ (a) and (b), in {3+,5+} combination there are three unpaired 
Ru electrons and four unpaired Co electrons alongwith two paired electrons while in {2+,6+} combination has two unpaired Ru electrons and three unpaired Co electrons along four paired electrons. \textcolor{black}{We consider the onsite level energy of Co t$_{2g}$ ($e_{Co}$), the t$_{2g}$-e$_g$ energy difference at Co site ($V_{Co}$), onsite level energy of Ru d ($e_{Ru}$), the intra-site 
Coulomb energy ($U_{Co3+}$, $U_{Co2+}$, $U_{Ru}$) and Hund's energy ($J_{Co}$, $J_{Ru}$). Note, we consider different onsite Hubbard $U$ values for different oxidation states of Co site. As Co is a 3d element,
the correlation effect is very prominent for smaller spatial
extension of 3d orbitals, which results significantly different $U$ values for different occupancies of Co-d levels. For higher oxidation number, the $U$ value becomes higher due to more compressive 3$d$ orbital. The values of the other parameters like $U_{Ru}$ for 4d Ru, or $J$ values do not change significantly between two different oxidation states. 
The energetics of the (3+,5+) and (2+,6+) thus can be written as, }

\textcolor{black}{
\begin{equation}
\begin{split}
E(3+, 5+)  &= 6e_{Co} + 2V_{Co} + 15U_{Co^{3+}} - 10 J_{Co} \\
 &+  3e_{Ru} + 3U_{Ru} - 3J_{Ru}\\
E(2+, 6+) &= 7e_{Co}+2V_{Co} + 21U_{Co^{2+}} - 11J_{Co} \\
&+ 2e_{Ru} + U_{Ru} - J_{Ru}\nonumber
\end{split}
\end{equation}
}
The energy difference ($\Delta E = E(3+, 5+) - E(2+, 6+)$) is given by
\textcolor{black}{
\begin{equation}
    \begin{split}
        \Delta E &= \Delta_{Co-Ru} + 15U_{Co^{3+}}-21U_{Co^{2+}}+ J_{Co}\\
        &+ 2U_{Ru} -2J_{Ru} \nonumber
    \end{split}
\end{equation}
}

$\Delta_{Co-Ru}$ = $e_{Ru} - e_{Co}$, estimated by NMTO downfolding calculation
considering an active basis of Co d and Ru d only states, turns out to be
$\sim$ 1.6 eV for monoclinic structure and 1.2 eV for tetragonal structure. \textcolor{black}{Following the first-principles estimates using linear response theory in Ref~\cite{co_u}, we have chosen $U_{Co^{3+}}\sim$ 6.7 eV and $U_{Co^{2+}}\sim$ 4.4 eV. For typical values of $U_{Ru}$, $J_{Co}$, and $J_{Ru}$ (U$_{Ru}$$\sim$1eV, J$_{Co}$$\sim$0.8 eV, J$_{Ru}$$\sim$0.5 eV)\cite{hubbardu_prb_22,hubbardu_prb_94,jh_prb_2011}, $\Delta E$ is found to be positive in the monoclinic phase, confirming stabilization of
(2+,6+) configuration. The stabilization of 
unusual valence of Ru 6+ in monoclinic CCRO is thus driven by the complex interplay of charge-transfer energy gain between Co and Ru, Hund energy gain at Co site, Coulomb energy gain at the Co site due to the decreased Hubbard $U$ in 2+ oxidation state,  and the Coulomb energy gain at the Ru site that over compensates the Hund energy loss at Ru site. This simplistic analysis, of course does not include the effect of structural relaxation,
hopping interactions etc, as in rigorous DFT calculation.}

\subsection{Mechanism of Magnetism and Magnetic Transition Temperature}

\begin{figure}
\centering
    \includegraphics[scale=0.6]{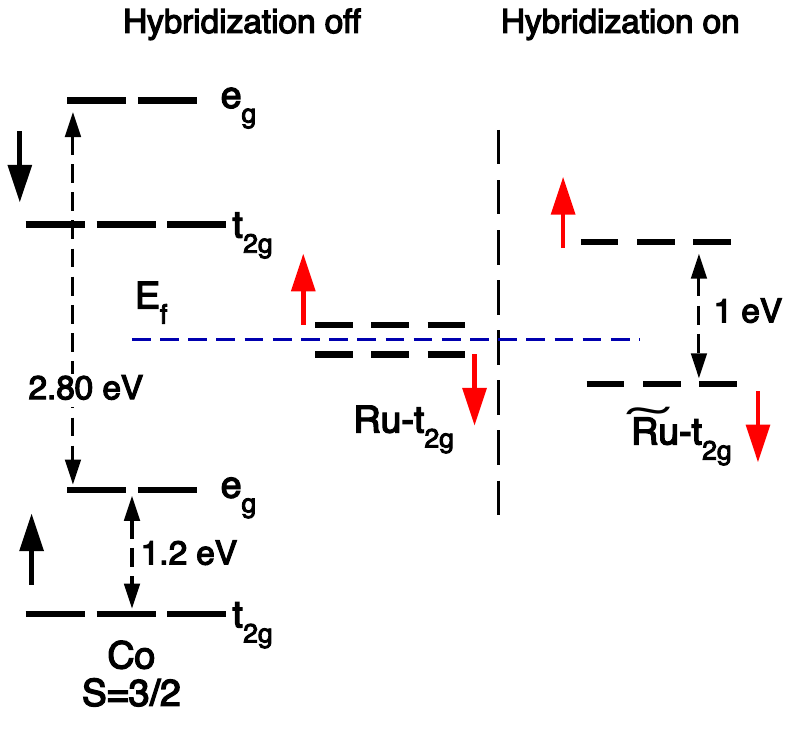}    
    \caption{The energy level positions of Co d and Ru t$_{2g}$ states, as given
    in NMTO-downfolding calculation considering Co d-Ru t$_{2g}$ basis (left) and
    massively downfolded Ru t$_{2g}$ only basis (right). The latter takes into
    account Co-d-Ru-t$_{2g}$ hybridization. The up and down spin channels of Co and Ru states are marked by black and red arrows respectively. The Fermi level is shown by blue dashed line.}
        \label{fig4:de}
\end{figure}

\begin{figure}[ht]
        \centering
	\includegraphics[width=0.95\linewidth, clip=true]{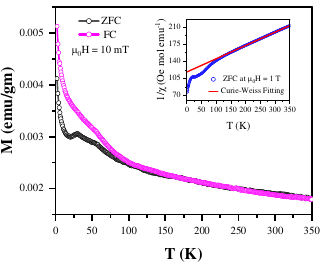}
	\caption{\textcolor{black}{Measured temperature-dependent magnetization. Inset shows the temperature dependence of inverse susceptibility.}}
	\label{fig-exp3}
\end{figure}

\begin{figure*}
\centering
    \includegraphics[scale=0.4]{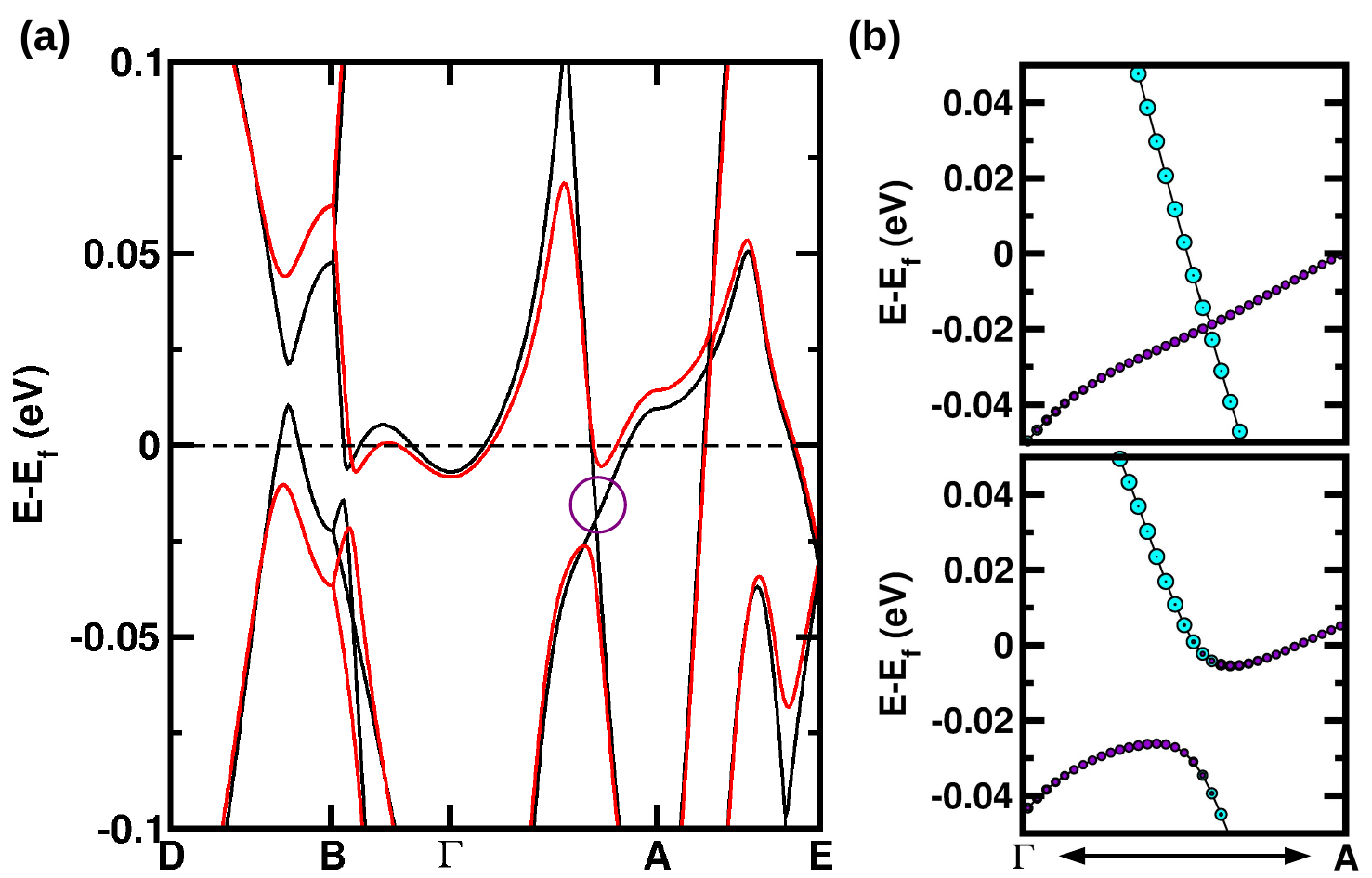}    
    \caption{(a) CCRO band structure in monoclinic crystal structure computed in 
    GGA+U (black) and GGA+U+SOC (red), plotted along the high symmetry point of the BZ, 
    D = (0.0 0.5 0.5), B = (0.0 0.0 0.5), $\Gamma$ = (0.0 0.0 0.0), A = (-0.5 0.0 0.5), and E = (-0.5 0.5 0.5). The Fermi level is set to zero.
    The anti-crossing point is encircled. (b) The zoomed band structure without (top) and with (bottom) SOC, projected to Ru-t$_{2g}$ orbital character. Ru $d_{xy}$, $d_{yz}$ and $d_{xz}$ characters are shown
    as brown, cyan and magenta colored circles, the radius of the circles being proportional to the weight of the orbital character.}
        \label{fig5:band}
\end{figure*}

The computed ground state of CCRO in the monoclinic structure is appealing which is half-metallic with Ru spins aligned antiparallel to Co spins. The situation
is akin to the celebrated case of Sr$_{2}$FeMoO$_6$ (SFMO), which in its half-metallic ground state shows the antiparallel alignment of Fe and Mo spins. The magnetism in SFMO has been explained by the unconventional two-sublattice double exchange mechanism, captured through strong coupling
between a core spin and an itinerant electron. In SFMO the Fe sites host the large, site-localized core spin S=5/2 and Mo sites provide the mobile electron that delocalizes on the Fe-Mo network. Coupling between the two strongly prefers one spin polarization of the itinerant electron, thus forcing Mo spins to align
antiparallel to core Fe spins. A similar scenario is expected to prevail
for CCRO. To explore this possibility, we performed calculations using the muffin-tin orbital-based technique of NMTO downfolding. NMTO downfolding calculations enable defining energy selected, effective Wannier functions by integrating out degrees of freedom that are not of interest (downfolding).
In the first step of this process, we downfolded all degrees of freedom except for Co-d and Ru-t$_{2g}$. The onsite block of the real space Hamiltonian obtained from this calculation provides information on the onsite energies of Co-d and Ru-t$_{2g}$, their intrinsic spin-splitting, and their relative alignment. In the second step, we performed a massive downfolding, keeping only Ru-t$_{2g}$ degrees of freedom active and downfolding all other degrees of freedom, including Co-d. The onsite block of the real space description resulting from this massive downfolding provides information on the energy levels of Ru-t$_{2g}$ and its spin splitting, which is induced by the hybridization between Co-d and Ru-t$_{2g}$. \textcolor{black}{Ru-t$_{2g}$ bands in the downfolded basis match well with that of entire DFT basis (see Section ~\ref{sec:suppsec6} and Fig. \ref{fig:nmto-dfld}}). Figure \ref{fig4:de} summarizes the positioning of the energy levels of CCRO in both the Co-d-Ru-t$_{2g}$ and massively downfolded, renormalized Ru-t$_{2g}$ basis, as determined by the NMTO downfolding calculations. Co-d states are crystal-field split (into t$_{2g}$ and e$_g$ manifolds) and spin split. The crystal field splitting estimated to be $\sim$ 1.2 eV is weaker compared to the significant spin splitting at the Co site ($\sim$ 2.8 eV), which arises from the large core spin of S = 3/2. The crystal field splitting of Ru 
is large due to the extended nature of 4d orbitals, thus only Ru t$_{2g}$ states are relevant. In Fig.\ref{fig4:de}, for simplicity, we show the energy positions of Co-t$_{2g}$, Co-e$_{g}$ and Ru-t$_{2g}$, neglecting the small splitting within each manifolds. In the Co-d-Ru-t$_{2g}$ basis, the Ru-t$_{2g}$ states are essentially nonmagnetic, exhibiting a small spin splitting of 0.2 eV. These states fall within the energy window of the strongly spin split states of Co-d  (cf left part of Figure \ref{fig4:de}). In the massively downfolded Ru-t$_{2g}$ basis, which accounts for the hybridization between Co-d and Ru-t$_{2g}$, 
an induced spin-splitting of approximately 1 eV develops at the Ru sites (left part of Figure \ref{fig4:de}), by pushing the up-spin Ru-t$_{2g}$ energy levels higher and the down-spin Ru-t$_{2g}$ energy levels lower. This induced spin splitting at Ru
is oppositely oriented to that of the Co-d. The mechanism of magnetism of CCRO
is thus governed by two sublattice double exchange mechanism which forces the induced moment at the Ru site to be opposite to that at the Co site, in turn stabilizing a ferromagnetic
ground state with parallel alignment of Co core spins. The two-sublattice double exchange mechanism is generally found to give rise to high magnetic transition temperatures, being dictated by the strength of B-B$ {'}$ hybridization.

The model Hamiltonian representation of the two-sublattice double exchange in CCRO, following the literature on SFMO\cite{sanyal_prb_2009}, can be written as,

\begin{eqnarray}
H&=&\epsilon_{Co}\sum_{i\in B}c_{i\sigma,\alpha}^{\dagger}c_{i\sigma,\alpha}+\epsilon_{Ru}\sum_{i\in B'}r_{i\sigma,\alpha}^{\dagger}r_{i\sigma,\alpha} \nonumber  \\
&-&t_{CR}\sum_{<ij>\sigma,\alpha}c_{i\sigma,\alpha}^{\dagger}r_{j\sigma,\alpha}-t_{RR}\sum_{<ij>\sigma,\alpha}r_{i\sigma,\alpha}^{\dagger}r_{j\sigma,\alpha} \nonumber \\
&-&t_{CC}\sum_{<ij>\sigma,\alpha}c_{i\sigma,\alpha}^{\dagger}c_{j\sigma,\alpha} \nonumber \\ &+ &J\sum_{i\in B}\vec{S}_{i}\cdot c_{i\alpha}^{\dagger}\vec{\sigma}_{\alpha\beta}c_{i\beta} + h.c.
\label{eqn1}
\end{eqnarray}

The $c$'s and the $r$'s refer to the Co and Ru sites, respectively. $t_{CR}$, $t_{RR}$, $t_{CC}$ represent the nearest-neighbour Co-Ru, second-nearest-neighbor Ru-Ru and Co-Co hoppings, respectively, the largest hopping being given by $t_{CR}$.  $\sigma$ is the spin index and $\alpha$ is the orbital index. 
The ${\vec S}_i$ is the `classical core spins (large $S$) at the Co site, coupled to itinerant Ru electrons through a coupling $J \gg t_{FM}$. \textcolor{black}{The $J$ term essentially represents a four-fermion interaction. While the core spin at the Co site is treated as a classical spin, it still retains full statistical dynamics and interacts with the itinerant electrons at the same site. This interaction gives rise to description of a correlated system. Similar approaches have been employed in many others previous studies ~\cite{Kumar2005,prb_raderia}. The resulting Hamiltonian is commonly known as the two-sublattice Kondo lattice model.}

To solve the Hamiltonian above, as in \cite{sanyal_prb_2009}, we assume
$\vert$ J $\vert$ $\to$ $\infty$ since $\vert$ J $\vert$ $\gg$ $\vert t \vert$. This results in an effective Hamiltonian, with "spinless" Co electrons, described by $\tilde{c}$, and Ru electrons having both spin states, with effective hopping picking up a ($\theta_i$, $\phi_i$) modulation, described by \\

$t_{CR}\sum_{<ij>}\left(\tilde{c}_{i}^{\dagger}r_{j\uparrow}sin\frac{\theta_{i}}{2}-\tilde{c}_{i}^{\dagger}r_{j\downarrow}cos\frac{\theta_{i}}{2}e^{i\phi}\right)$

($\theta_i$, $\phi_i$) describes the spin axis of the Co core spin. For example, $\theta$ = 0, $\phi$= 0, corresponds to FM configuration with all Cr core spins being up. Since the Co core spin, S=3/2 is large and can be considered classical, one can consider different spin configurations (FM, AFM, and disordered) and diagonalize the system in real space to obtain variational estimates of the ground state and its stability. The calculations were carried out for finite-size  lattices of dimensions
8 $\times$ 8 $\times$ 8. The hopping parameters and on-site energies were taken from the DFT calculations. We considered two different spin arrangements of the Co core spins, FM and G-type AFM, and measured their energies with respect to a spin-disordered phase, mimicking the paramagnetic (PM) phase. The PM energy is obtained by averaging the energy of 50 such configurations. The stability of the FM arrangements of Co spins can be measured as the energy difference ($\Delta$E) between the PM and FM phases. The calculated $\Delta$E for monoclinic CCRO is found to be $\sim$ 120 K. 

\textcolor{black}{As mentioned earlier, all different synthesis attempts, resulted in samples containing a fraction of impurity phase, which complicates the magnetic characterization.  In particular, the impurity phase Ca$_3$Co$_4$O$_9$ is known to exhibit multiple magnetic transitions, including a ferrimagnetic transition below 19K~\cite{Sugiyama2003}. Despite this difficulty, to have an experimental support of our theoretically predicted magnetic transition, we performed temperature-dependent magnetization measurements. As shown in Fig.~\ref{fig-exp3}, the magnetization $M(T)$ curve exhibits a bifurcation between the zero-field-cooled (ZFC) and field-cooled (FC) branches near 100K, suggesting the onset of a magnetic transition. The bifurcation temperature is within acceptable range of theoretically calculated temperature considering both theoretical approximations and the influence of sample purity.}

\begin{figure}
\centering
    \includegraphics[scale=0.65]{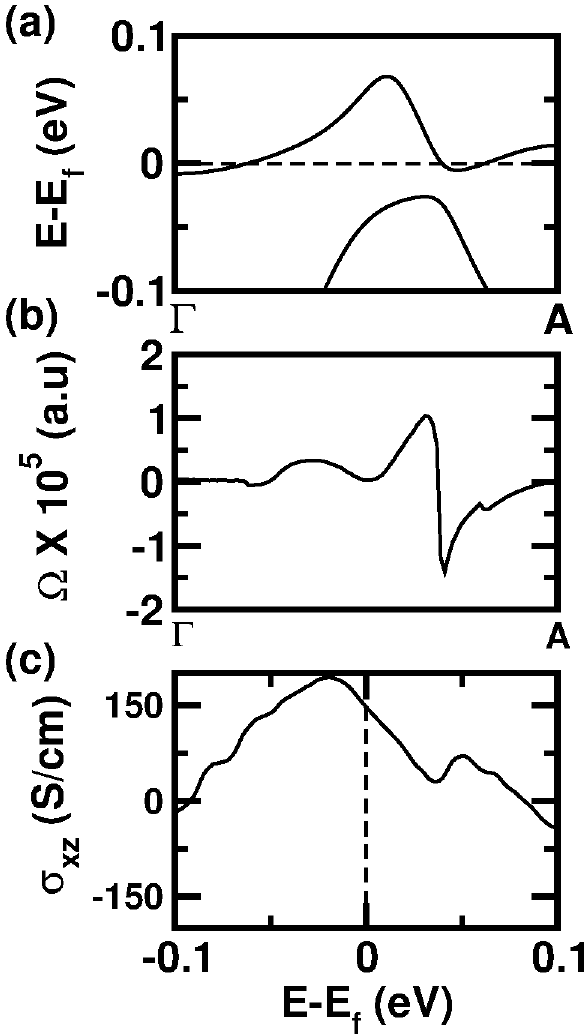}    
    \caption{(a) GGA + U+ SOC band structure of monoclinic CCRO along $\Gamma$ to A within 100 meV along Fermi level. (b) Berry curvature corresponding to the anti-crossing bands, along $\Gamma$ to A. (c) The anomalous Hall conductivity, plotted as a function of energy.}
        \label{fig6:ahc}
\end{figure}

\subsection{Topological Properties - Anomalous Hall Conductivity}

 The above analysis establishes that CCRO in ground state monoclinic structure is metallic with a net magnetic moment of 1 $\mu_B$/f.u. Due to the partially filled Ru-t$_{2g}$ states, the spin-orbit coupling effect is also expected to be non-negligible. This motivates us to explore the anomalous Hall conductivity in this compound. For this purpose, we first examine the band structure with and without SOC. Fig-$\ref{fig5:band}$ shows the band structures and the orbital projected band structures of CCRO, plotted around a narrow energy window of 100 meV around the Fermi energy (E$_f$) along the high symmetry point of the monoclinic Brillouin Zone (BZ). In the GGA+U band structure, we find the presence of low-energy Dirac-like crossing (encircled in Fig-$\ref{fig5:band}$(a)) between $\Gamma$ and A high symmetry points at {\bf K$_c$} = (-0.34 0.0 0.34), lying about 15 meV below E$_f$. Switching on the SOC, in the GGA+U+SOC band structure, 
 a gap opens up at {\bf K$_c$}. As observed from the orbital character projected band structures (cf. Fig-$\ref{fig5:band}$(b)) the crossing at {\bf K$_c$} arises due to the intersection of Ru t$_{2g}$ bands of two different orbital characters, Ru yz and xz. Upon inclusion of SOC, a band anticrossing happens, resulting in a change of the band orbital character of the two bands involved in forming the crossing, as the k vector changes from a smaller value to a value larger than that of {\bf K$_c$} \cite{anticrossing}. As discussed in the context of materials like LaBi \cite{lou2017evidence,niu2016presence}, such a SOC-driven anticrossing can open up a topological gap. To confirm the nontriviality of the anticrossing points, we then calculated the contribution of the Berry curvature for the pair of bands involved in the anticrossing, using Eqn.\ref{eqn:berry}. Integration of the Berry curvature arising
from the bands forming the anticrossing over the K2=0 2D plane (cf Eqn.\ref{eqn:chern}) resulted in a Chern number with an integer value of 1.  K1-K3 plane was chosen since the anti-crossing point {\bf K$_c$} is found to lie in this plane. This establishes that the pair of bands involved in the anticrossing exhibit nontrivial topology. It is important to note that the Chern number, C, is well-defined in the insulating state, taking integer values. The Chern number in this context is calculated for a specific pair of bands forming the anticrossing, which remains well-separated at all momenta. There are other trivial bands which cross E$_f$ along with one of these
two separated bands to make CCRO a Chern metal. The anti-crossing point at {\bf K$_c$}  thus acts as a magnetic monopole giving rise to a fictitious magnetic field, responsible for the anomalous Hall conductivity. The band dispersion along $\Gamma$-A direction, along with the corresponding Berry curvature, computed for the pair of bands forming the anticrossing, is presented in Fig-\ref{fig6:ahc}(a) and (b). Since these bands also intersect with other trivial bands, the calculation was extended to include the entire subset of bands they cross. The contribution of these bands to the Berry curvature was found to be minimal. In particular, their effect was found to be negligible at the anticrossing point, where the Berry curvature peaks. The calculated intrinsic AHC, $\sigma_{xz}$, obtained by integrating the Berry curvature over k-space and considering the bands up to energy E (cf Eqn.\ref{eqn:ahc}), is shown in Fig-\ref{fig6:ahc} (c). The value of AHC at E$_f$ turns out to be 147 S/cm.  It is worth noting that $\sigma_{xz}$ peaks at an energy corresponding to the anti-crossing point. This leads to the conclusion that the anti-crossing point contributes to the Berry curvature leading to nontrivial topology and AHC. \textcolor{black}{The computed value of AHC for CCRO is comparable to that predicted recently for a set of 3d-4d/5d double perovskites\cite{samanta_npj_2023large}. CCRO thus adds to the expanding material class of compounds, exhibiting intrinsic AHC 
arising out of Berry curvature effect.}

\section{Summary and Discussion}

Keeping in mind the importance of double perovskite compounds with a combination of 3d TM and 4d or 5d TM at B and B$^{'}$ sites, in this study we report the synthesis of two Ru-based ordered double perovskites, Ca$_2$FeRuO$_6$ and Ca$_2$CoRuO$_6$. To the best of our knowledge, these compounds have not been
synthesized before, though a disordered form of  Ca$_2$FeRuO$_6$ has been reported earlier\cite{Naveen2018}. Refinement of the crystal structure of Ca$_2$FeRuO$_6$ and Ca$_2$CoRuO$_6$ led to monoclinic P2$_1$/c symmetry for both compounds. The monoclinic symmetry of the synthesized CFRO compound signals the ordering of Fe and Ru, as opposed to the previously reported orthorhombic symmetry of CFRO with disordered Fe/Ru sites.\cite{Naveen2018} The ordering between Fe and Ru sites in the present study was achieved by lowering the growth temperature. Interestingly while the experimentally determined crystal symmetry matches with ML-high throughput predicted symmetry for CFRO, for CCRO the ML-high throughput predicted symmetry was found to be tetragonal I4/m. Accurate first-principles calculation confirmed the stability of monoclinic P2$_1$/c symmetry
over the predicted tetragonal I4/m symmetry for CCRO. \textcolor{black}{
The change in symmetry from tetragonal to monoclinic influences the electronic structure of CCRO in a non-trivial manner. While CCRO in tetragonal symmetry hosts 3+/5+ nominal valence of Co/Ru, in the ground state monoclinic symmetry CCRO is found to host unusual 6+ valence of Ru
together with 2+ valence of Co. The theoretically predicted 6+ valence
of Ru has been confirmed through BVS analysis of experimentally refined
structure.} The high oxidation state of 6+ of Ru 
in octahedral coordination is not reported, though it is reported in tetrahedral coordination \cite{k2ruo4}. \textcolor{black}{The first-principles calculated band structure of the monoclinic CCRO turns out
to be half-metallic with magnetism driven by two-sublattice double exchange mechanism involving the localized Co S=3/2 spin and itinerant Ru electrons strongly hybridized with Co minorty spin electrons, akin
to that of SFMO \cite{sfmp_prl_sharma}.} The calculated T$_c$ via
the exact diagonalization of the DFT-derived model Hamiltonian turns out to be $\sim$ 120 K, \textcolor{black}{which is in line with measured temperature-dependent magnetization data.} 

\textcolor{black}{Inclusion of SOC in the calculated band structure
shows nontrivial band anticrossing between Ru $d_{yz}$ and $d_{xz}$ bands is found close to E$_f$ ($\sim$ 15 meV below E$_f$). This generates a fictitious magnetic field, which is responsible for the intrinsic anomalous Hall conductivity. The AHC arising from the Berry curvature is found to be 147 S / cm, greater than that calculated for the well-studied double perovskite compound SFMO, and comparable to that calculated for Sr$_2$NiOsO$_6$ \cite{samanta_npj_2023large}.}

\textcolor{black}{Finally, our theoretical calculations did not consider the effect of anti-site disorder present in the sample. This issue has been investigated extensively in Ref.\cite{ahalder_prb_disod}. As is explicitly shown in this reference, the magnetic behavior under two site double exchange is robust against anti-site disorder. When magnetism is governed by the double exchange mechanism - as is the case in our system - the ordered magnetic moment as well as the magnetic transition temperature are largely retained, even in presence of anti-site disorder. The anti-site disorder, on the other hand, may affect the transport properties, importantly the half-metallic behavior. As presented in section ~\ref{sec:suppsec4}, the experimental transport data support a narrow band gap of 85 meV, in comparison to half-metallic character of theoretical DOS. In particular, the dip observed in the minority spin DOS (cf Fig. 4(b)) may get 
localized due to anti-site disorder.}

\textcolor{black}{In closing, we underline the important contribution of experiment in discovering the correct space group symmetry. This is found to be instrumental in stabilizing the 6+ valence of Ru and thus influencing the calculated physical properties. Considering the synthesis and characterization challenges, as detailed above, our theoretical results 
provide the guiding insights which should motivate further
experimental investigations.}

\section{Supplementary Information}{\label{sec:SI}}

\subsection{\texorpdfstring{Synthesis Attempts and Phase Purity Challenges in $Ca_2CoRuO_6$}{Synthesis Attempts and Phase Purity Challenges in Ca_2CoRuO_6}}\label{sec:suppsec1}

\begin{figure*}
    \centering
    \includegraphics[scale=0.8]{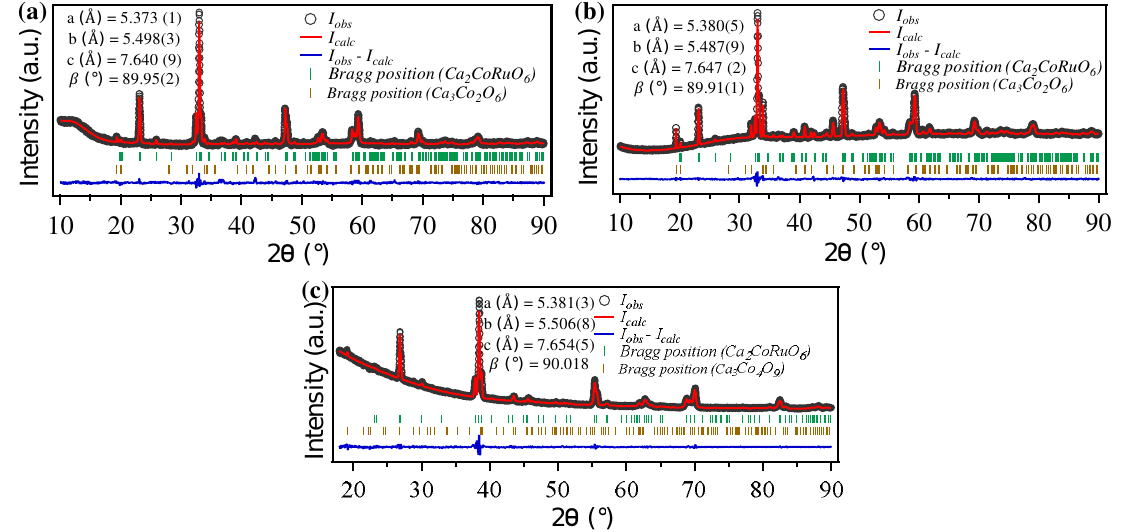}
    \caption{(a) XRD pattern with Rietveld refinement for solid-state Synthesized Sample. (b) XRD pattern with Le Bail fitting for sol-gel synthesized sample. (c) XRD pattern with Le Bail fitting for high-pressure O$_2$ synthesized sample}
    \label{fig1}
\end{figure*}

To synthesize Ca$_2$CoRuO$_6$, we employed three distinct synthesis routes: (i) solid-state reaction in ambient air, (ii) sol-gel method in ambient air, and (iii) solid-state reaction under a high-pressure oxygen atmosphere. Despite varying the synthesis conditions, powder X-ray diffraction (XRD) analysis revealed the presence of secondary phases in all samples, as shown in Fig.~\ref{fig1}.

In our initial attempt, Ca$_2$CoRuO$_6$ was synthesized using a conventional solid-state reaction route. Stoichiometric amounts of CaCO$_3$, Co$_3$O$_4$, and RuO$_2$ were thoroughly mixed in an agate mortar for approximately two hours to ensure homogeneity. The resulting mixture was calcined at 600$^\circ$C for 24 hours to mitigate Ru volatility, followed by sintering at 1000$^\circ$C for 48 hours in air. The sample was subsequently cooled at a rate of 100$^\circ$C per hour. Rietveld refinement of the XRD pattern revealed the presence of a secondary phase, Ca$_3$Co$_2$O$_6$, coexisting with the target compound as shown in Fig.~\ref{fig1}(a). Notably, Ca$_3$Co$_2$O$_6$ is known to undergo complex magnetic transitions, including a partially disordered antiferromagnetic ordering below 25~K~\cite{PhysRevLett.101.097207}, complicating the interpretation of the intrinsic magnetic behavior of Ca$_2$CoRuO$_6$.

In an effort to improve phase purity, we then employed a sol-gel synthesis route. For this, CaCO$_3$ was converted to Ca(NO$_3$)$_2$ using concentrated nitric acid. The precursor solution was prepared using Ca(NO$_3$)$_2$, Co(NO$_3$)$_2\cdot$6H$_2$O, and RuO$_2$ along with citric acid (C$_6$H$_8$O$_7$) and ethylene glycol (C$_2$H$_6$O$_2$) as complexing agents and distilled water as the solvent. The dried gel was decomposed and calcined at various temperatures, with 1050$^\circ$C identified as optimal for phase formation. However, XRD analysis using Le Bail fitting still showed the persistence of Ca$_3$Co$_2$O$_6$ as a prominent impurity phase as shown in Fig.~\ref{fig1}(b).

To further suppress impurity phases and promote better cation ordering, we performed solid-state synthesis under high-pressure oxygen conditions, as detailed in the main manuscript. This approach, however, yielded a new impurity phase, Ca$_3$Co$_4$O$_9$, according to the Le Bail fitting on the XRD pattern as shown in Fig.~\ref{fig1}(c). Ca$_3$Co$_4$O$_9$ also exhibits magnetic ordering with multiple transitions, including a ferrimagnetic transition below 19~K~\cite{PhysRevB.67.104410}, which again complicates the interpretation of the magnetic measurements for the target compound.

These synthesis trials demonstrate the significant challenge in obtaining a phase-pure sample of Ca$_2$CoRuO$_6$, particularly due to the competing formation of thermodynamically stable cobalt-rich phases. Ca$_3$Co$_4$O$_9$ is thermally stable up to approximately 926$^\circ$C, above which it decomposes into Ca$_3$Co$_2$O$_6$, CoO, and CaO. Upon further heating beyond $\sim$1026$^\circ$C, Ca$_3$Co$_2$O$_6$ itself becomes unstable and breaks down into CaO and CoO~\cite{Wongdamnern2022}. Future synthesis efforts will focus on fine-tuning oxygen partial pressure and reaction kinetics to suppress impurity formation and reduce antisite disorder.

\begin{figure}
    \centering
    \includegraphics[width=0.4\textwidth]{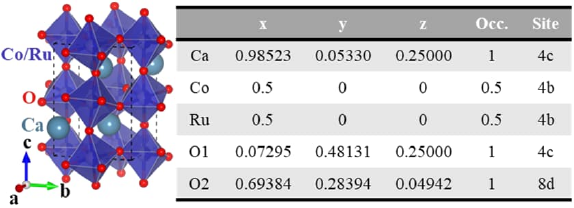}
    \caption{Structural model of Ca$_2$CoRuO$_6$ in orthorhombic (Pbnm No. 62) phase.}
    \label{fig2}
\end{figure}

\begin{figure*}
    \centering
    \includegraphics[width=0.5\textwidth]{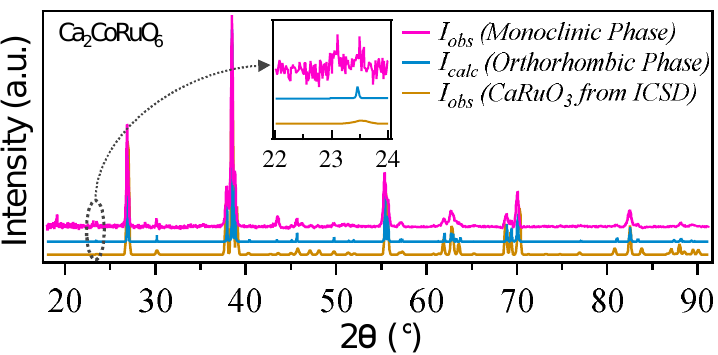}
    \caption{Comparison of experimental XRD with that of an orthorhombic (Pbnm No. 62) structure and CaRuO$_3$ XRD pattern.}
    \label{fig3}
\end{figure*}

\subsection{\texorpdfstring{Structural characterization on $Ca_2CoRuO_6$}{Structural characterization on Ca_2CoRuO_6}}\label{sec:suppsec2}

To confirm the rock-salt-type ordering of Co and Ru cations in Ca$_2$CoRuO$_6$, thorough XRD analysis were performed. Rietveld refinement using the monoclinic $P2_1/n$ space group reveals the presence of two distinct reflections at $2\theta = 23.07^\circ$ and $23.43^\circ$, which correspond to the (011) and (101) crystallographic planes, respectively. These so-called R-point reflections are characteristic signatures of rock-salt-type ordering at the B-site in double perovskite structures.

In contrast, a structural model based on an orthorhombic symmetry [as illustrated in Fig.~\ref{fig2}] displays only a single reflection at $2\theta = 23.43^\circ$, corresponding to the (101) plane. To emphasize the significance of the double-peak feature in this low-angle region, we compared the experimental XRD pattern of Ca$_2$CoRuO$_6$ (modeled in the monoclinic phase) with that of the orthorhombic model and with a reference pattern of the GdFeO$_3$-type disorder perovskite structure CaRuO$_3$, obtained from the ICSD database~\cite{Bensch1990}. The CaRuO$_3$ pattern shows only one peak at $2\theta = 23.52^\circ$ with comparatively weak normalized intensity, as illustrated in Fig.~\ref{fig3}. This comparative analysis supports the conclusion that our synthesized Ca$_2$CoRuO$_6$ sample crystallizes in the monoclinic $P2_1/n$ phase with Co and Ru exhibiting rock-salt-type ordering.

\begin{figure*}
    \centering
    \includegraphics[width=0.8\textwidth]{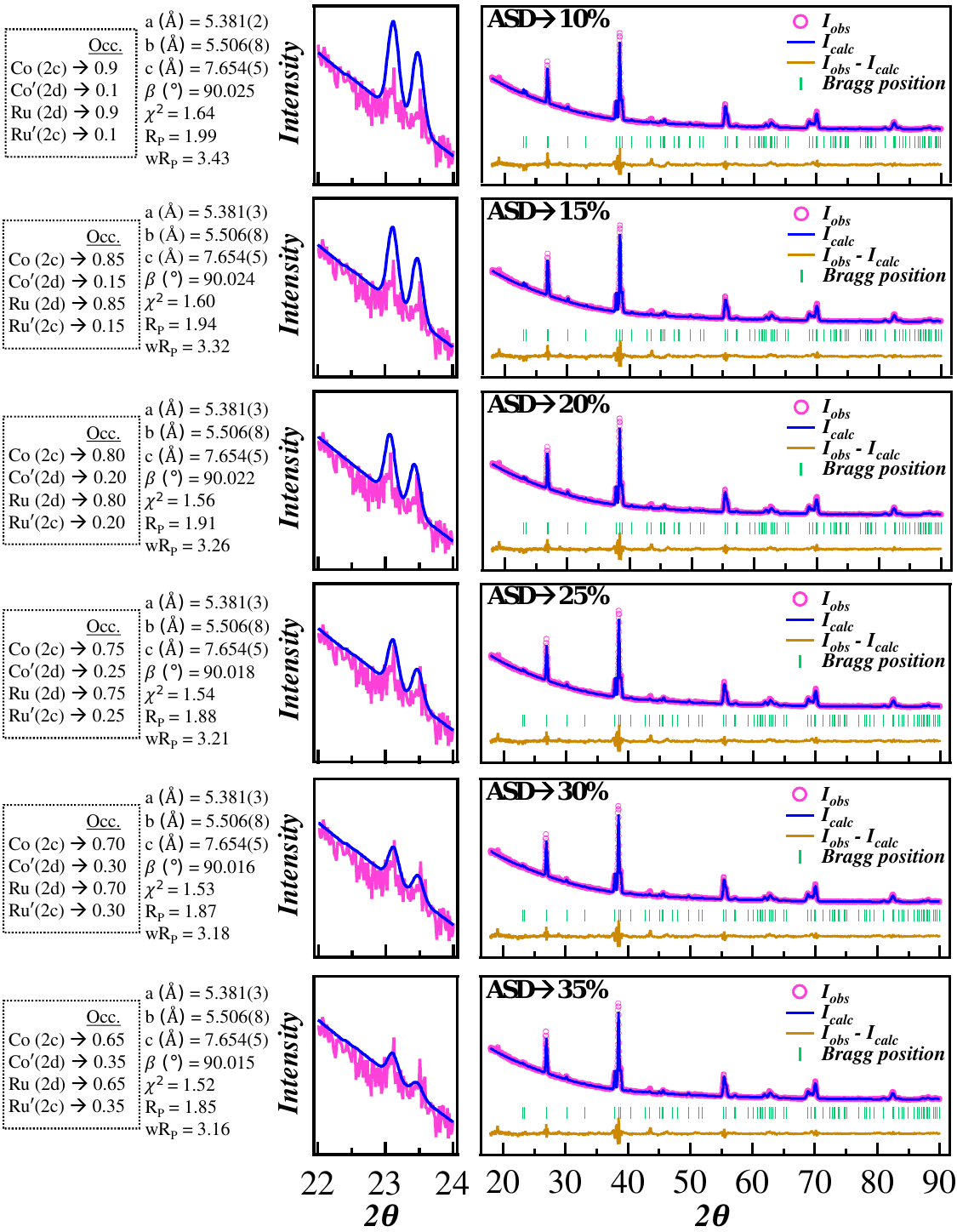}
    \caption{Rietveld refinement using the monoclinic $P2_1/n$ structure by varying the antistites disorder for Ca$_2$CoRuO$_6$.}
    \label{fig4}
\end{figure*}

To further refine the structural model, we analyzed the influence of antisite disorder as the best fitting of the low-angle double peak structure in the experimental XRD pattern of Ca$_2$CoRuO$_6$. By systematically varying the site occupancies of Co and Ru at the 2$c$ and 2$d$ Wyckoff positions within the monoclinic $P2_1/n$ symmetry, we found that a model incorporating $\sim$25\% antisite disorder provides the best fit to the observed double-peak structure. Fig.~\ref{fig4} shows the comparative fits, confirming the presence of partial disorder between Co and Ru ions at the B-sites in Ca$_2$CoRuO$_6$.

\subsection{\texorpdfstring{Bond Valence Sum analysis on $Ca_2CoRuO_6$}{Bond Valence Sum analysis on Ca_2CoRuO_6}}\label{sec:suppsec3}
To estimate the oxidation states of the transition metal cations in Ca$_2$CoRuO$_6$, we performed a bond valence sum (BVS) analysis using the refined Co-O and Ru-
O bond lengths obtained from Rietveld refinement of the X-ray diffraction data. The BVS method provides an empirical estimation of the valence of a cation based on its bonding environment.

The valence $V$ of a cation is calculated using the following expression:
\[
V = \sum_j \exp\left(\frac{R_0 - R_{ij}}{B}\right)
\]
where $R_{ij}$ is the bond length between the cation and the $j$-th coordinated anion, $R_0$ is the tabulated bond valence parameter specific to a given cation-anion pair, and $B$ is an empirical constant typically taken to be 0.37~\AA. The summation runs over all anion neighbours, yielding an estimate of the effective oxidation state of the cation. For example, when applying the BVS method using the bond valence parameters corresponding to a +2 oxidation state, the resulting valence sum for a metal ion will typically approach an integer value of two. In contrast, using parameters associated with other oxidation states may also yield a value near two, but with a noticeably larger deviation from the nominal oxidation state, indicating a poorer match to the actual oxidation state~\cite{Nguyen2020}.

For the BVS calculations, we utilized the Co-O and Ru-O bond lengths obtained from Rietveld refinement of the monoclinic structure of Ca$_2$CoRuO$_6$, as reported in Table I of the main manuscript. The resulting bond valence sums, along with the corresponding absolute deviations from the nominal oxidation states for both Co and Ru, are presented in Table~\ref{tab1}.

\begin{table}
\caption{\label{tab1}Summary of the bond valence sum analysis.}
\begin{ruledtabular}
\begin{tabular}{lccc}
\textrm{Cation} & \textrm{Nominal Valence} & \textrm{Calculated BVS} & \textrm{Valence Difference}\\
 & & &\textrm{$(|Nominal - BVS|)$} \\
\colrule
Co$^{2+}$ & 2.00 & 2.48 & 0.48 \\
Co$^{3+}$ & 3.00 & 2.14 & 0.86 \\
Ru$^{6+}$ & 6.00 & 5.55 & 0.45 \\
Ru$^{5+}$ & 5.00 & 6.03 & 1.03 \\
\end{tabular}
\end{ruledtabular}
\end{table}

As observed, the calculated BVS for Co$^{2+}$ is 2.48, showing a smaller deviation (0.48) compared to that for Co$^{3+}$ (0.86), while the BVS for Ru$^{6+}$ is 5.55 with a deviation of 0.45, much closer than the value for Ru$^{5+}$ (6.03, deviation 1.03). These results support the assignment of Co$^{2+}$ and Ru$^{6+}$ oxidation states in the Ca$_2$CoRuO$_6$ structure, in agreement with our theoretical calculations.

\subsection{\texorpdfstring{Resistivity Measurements on $Ca_2CoRuO_6$}{Resistivity Measurements on Ca_2CoRuO_6}}{\label{sec:suppsec4}}

\begin{figure*}
    \centering
    \includegraphics[width=0.5\textwidth]{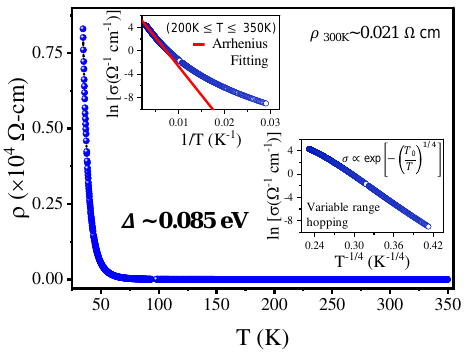}
    \caption{Temperature-dependent resistivity data of Ca$_2$CoRuO$_6$.}
    \label{fig5}
\end{figure*}

We also performed \textcolor{black}{direct current four-probe} resistivity measurements \textcolor{black}{using ETO option of a Quantum Design physical property measurement system (PPMS) without any applied magnetic field}. The results are shown in Fig.~\ref{fig5}, which reveal semiconducting behaviour.
In the high-temperature regime (200-350 K), the temperature dependence of electrical conductivity $\sigma = 1/\rho$ can be well described by an Arrhenius-type activated transport model,
\[
    \sigma(T) = A \exp\left(-\frac{E_a}{k_B T}\right),
\]
where $A$ is a material-dependent prefactor, $E_a$ is the activation energy, and $k_B$ is the Boltzmann constant. The linearity of $\ln(\sigma)$ versus $1/T$ in this regime supports thermally activated conduction with an activation energy $E_a \approx 0.085$ eV, as extracted from the fit shown in the top-left inset of Fig.~\ref{fig5}.
At lower temperatures ($<$250 K), the resistivity deviates from the Arrhenius model, suggesting the dominance of a different conduction mechanism. In this 
regime, the transport is better described by Mott's three-dimensional variable-range hopping (VRH) model, which is typically observed in disordered or localized 
systems. In VRH conduction, charge carriers hop between localized states and the conductivity is given by,
\[
    \sigma(T) = \sigma_0 \exp\left[-\left(\frac{T_0}{T}\right)^{1/4}\right],
\]
where $\sigma_0$ is the conductivity prefactor and $T_0$ is a characteristic temperature. As shown in the bottom-right inset of Fig.~\ref{fig5}, the plot of 
$\ln(\sigma)$ versus $T^{-1/4}$ displays a linear behavior in the low-temperature range, confirming that VRH dominates the charge transport mechanism below 250 K. Therefore,
this small but finite gap \textcolor{black}{of 85 meV} is inconsistent with the half-metallic behavior predicted theoretically. We attribute this deviation to the presence of antisite disorder, which was estimated to be $\sim$25\% from Rietveld refinement of our X-ray diffraction data, as well as the presence of impurity phase. In double perovskites, such disorder disrupts the long-range ordering of B/B$'$ cations, introduces mid-gap impurity states, and reduces carrier mobility factors. In particular, the dip structure, found in the computed
minority spin density of state \textcolor{black}{(cf Fig. 4(b) of main text)} is prone to such effects. 
We anticipate that further optimization to produce cleaner, phase pure samples will enhance this response and provide a more direct probe of intrinsic bulk transport.

\subsection{NMTO Downfolding Calculation}\label{sec:suppsec6}
\begin{figure*}
    \centering
    \includegraphics[scale=0.5]{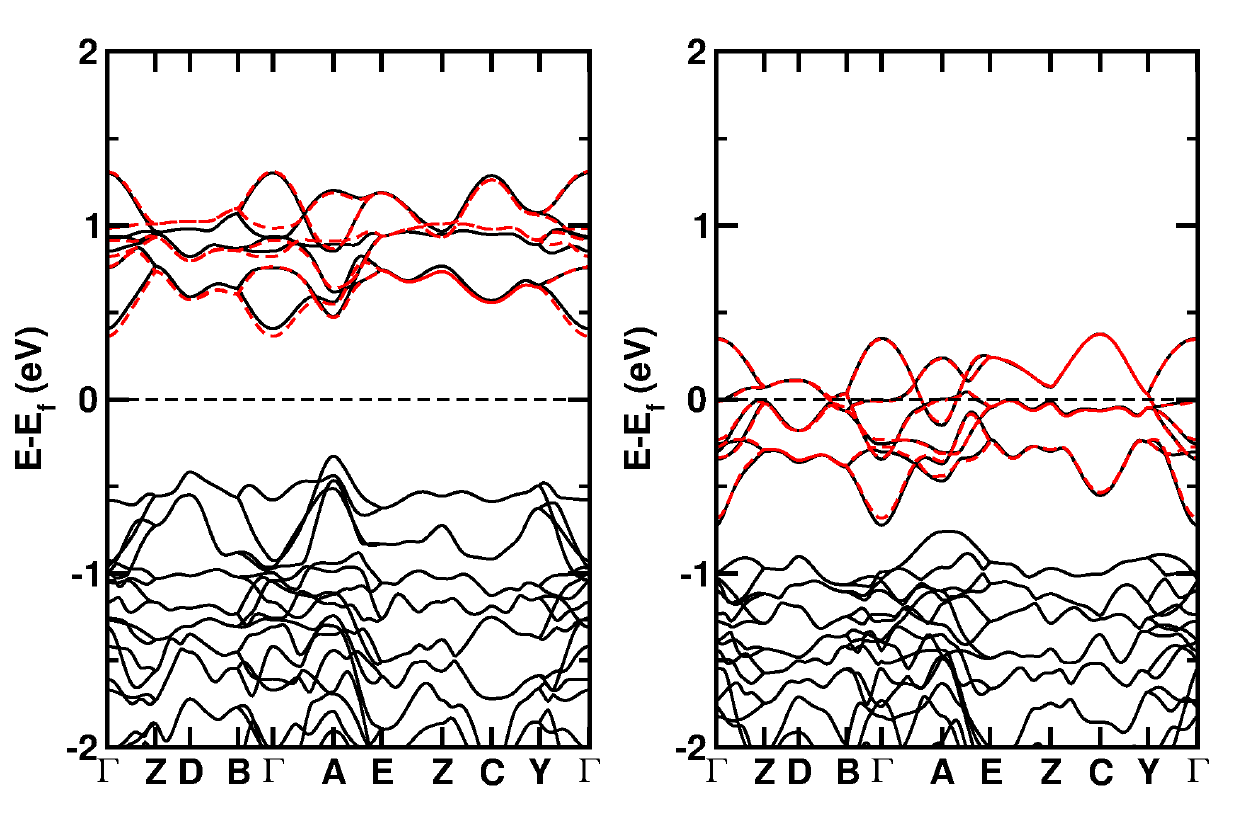}
    \caption{The left and right panels display the Ru-t$_{2g}$ bands in the downfolded basis alongside the full DFT band structures for the majority and minority spin channels, respectively. The black solid lines represent the band structures from the full DFT basis, while the red dashed line represents band structures in the downfolded basis.}
    \label{fig:nmto-dfld}
\end{figure*}

\begin{figure*}
    \centering
    \includegraphics[scale=0.4]{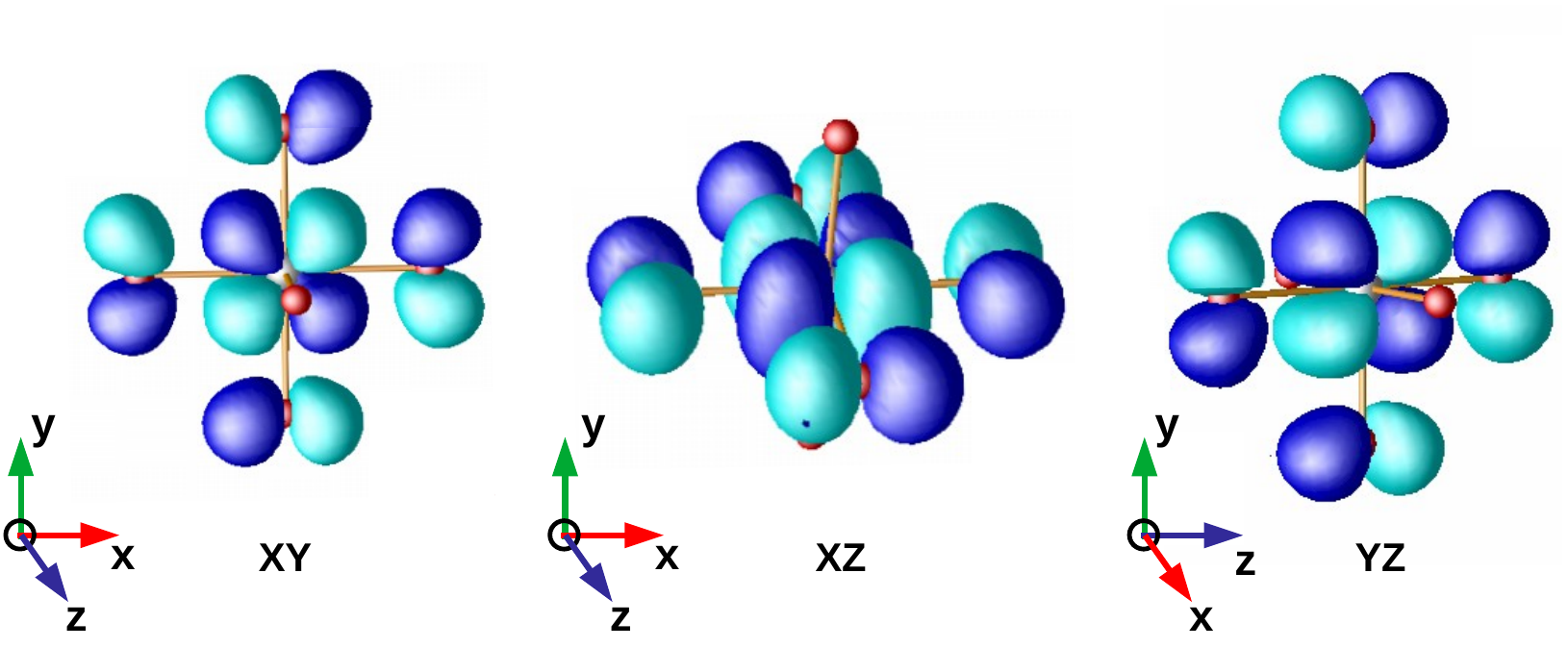}
    \caption{Ru-$t_{2g}$ ($xy$, $yz$, $xz$) Wannier functions obtained through NMTO downfolding calculations. Ru and O atoms are represented by grey and red spheres, respectively. Lobes of opposite signs are depicted in blue and cyan colors.}
    \label{fig:mlwf}
\end{figure*}

The low-energy effective Hamiltonian was constructed using the NMTO downfolding technique~\cite{nmto_prb_2000}, which integrates out irrelevant orbitals within a specific energy window and forms a renormalized basis. In the first step of our calculation, we downfolded all the degrees of freedom except for Co-d and Ru-t$_{2g}$ orbitals, to extract out their intrinsic spin splittings, and their relative alignment. In the second step, we carried out an extensive downfolding, retaining only the Ru-t$_{2g}$ orbitals while integrating out all others, including Co-d to calculate the  spin splitting of Ru-t$_{2g}$, which is induced
by the hybridization between Co-d and Ru-t$_{2g}$. Fig.~\ref{fig:nmto-dfld} shows the Ru-$t_{2g}$ bands derived from the downfolded model overlaid on the full DFT band structure. It is evident from Fig.~\ref{fig:nmto-dfld} that the Ru-$t_{2g}$ bands obtained through downfolding closely match with the full DFT band
structure. The maximally localized Wannier functions corresponding to the Ru-$t_{2g}$ orbitals are illustrated in Fig.~\ref{fig:mlwf}. As evident from the 
figure, the central part of the Wannier functions resembles the shape of the respective $t_{2g}$ orbital, while the tails extend toward neighboring O atoms, indicating significant $d$-$p$ $\pi$-type hybridization.

\newpage
\bibliography{ref}
\end{document}